\documentclass[12pt]{article}
\let\ssection=\section
\renewcommand{\section}{\setcounter{equation}{0}\ssection}

\newcommand\mathC{\mkern1mu\raise2.2pt\hbox{$\scriptscriptstyle|$}
        {\mkern-7mu\rm C}} % The complex numbers
\newcommand{\mathR}{{\rm I\! R}}         % The real numbers

\newcommand\bi{\begin{itemize}}
\newcommand\ei{\end{itemize}}
\newcommand\be{\begin{equation}}
\newcommand\ee{\end{equation}}

\newcommand{\dd}{\ensuremath{{\delta}}}
\newcommand{\al}{\ensuremath{{\alpha}}}

\newcommand{\pl}{\ensuremath{{\partial}}}

\begin{document}
\begin{titlepage}

\begin{center}
{\large\bf Some Aspects of Modality in Analytical Mechanics}
\end{center}

\vspace{0.8 truecm}

\begin{center}
        J.~Butterfield\footnote{email: jb56@cus.cam.ac.uk;
            jeremy.butterfield@all-souls.oxford.ac.uk}\\[10pt] All %%@
Souls College\\ Oxford OX1 4AL
\end{center}

\begin{center}
       20 Feb 2003\\
		Dedicated to the memory of David Lewis

\vspace{0.8 truecm}
		
		For the book, {\em Formal Teleology and Causality}, ed. M. %%@
St\"{o}ltzner, P. Weingartner, Paderborn: Mentis. 
\end{center}

\vspace{0.8 truecm}

\begin{abstract}
This paper discusses some of the modal involvements of analytical %%@
mechanics. I first review the elementary aspects of the Lagrangian, %%@
Hamiltonian and Hamilton-Jacobi approaches. I then discuss  two modal %%@
involvements; both are related to David Lewis' work on modality, %%@
especially on counterfactuals.\\
\indent The first is the way Hamilton-Jacobi theory uses ensembles, %%@
i.e. sets of possible initial conditions. The structure of this set of %%@
ensembles remains to be explored by philosophers.\\
\indent The second is the way the Lagrangian and Hamiltonian %%@
approaches' variational principles state the law of motion by %%@
mentioning contralegal dynamical evolutions. This threatens to %%@
contravene the principle that any actual truth, in particular an actual %%@
law, is made true by actual facts. Though this threat can be avoided, %%@
at least for simple mechanical systems, it repays scrutiny; not least %%@
because it leads to some open questions. 
 \end{abstract}

\end{titlepage}

\section{Introduction}\label{Intrpsa}
Ever since its beginnings, analytical mechanics has been a rich field %%@
for philosophical exploration. In particular, the principle of least %%@
action---with its various forms, and its strong suggestion of %%@
teleology---has been a focus of discussion from the time of Maupertuis %%@
to now: as witness some of the essays in this volume, and other %%@
excellent recent work such as St\"{o}ltzner (2003). However, so far as %%@
I can tell, philosophers have not explored the modal involvements of %%@
analytical mechanics. So I propose in this paper to make a first foray %%@
into this territory.

More specifically, I will discuss two modal involvements. Both are %%@
related to David Lewis' work on modality, especially his theory of %%@
counterfactuals (1973). (The first concerns Hamilton-Jacobi theory; the %%@
second Lagrangian and Hamiltonian mechanics.) So I dedicate the paper %%@
to his memory. Although analytical mechanics was not a topic central to %%@
his interests, the discussion will illustrate a view central to his %%@
metaphysical system, and to his influence on analytical philosophy: %%@
that science, indeed all our knowledge and belief, is steeped in %%@
modality. Besides, any philosopher who knew Lewis the man as well as %%@
the work, knows not only that he was a  
great philosopher---with transcendent creativity and craftsmanship, and %%@
enormous intellectual generosity---but also that he had wide %%@
intellectual interests in the sciences. So I like to think my %%@
illustrations of Lewisian themes in mechanics would have pleased him.

The modal involvements of analytical mechanics turn out to be rich and %%@
subtle. There is much  to explore here: as so often in the philosophy %%@
of physics, one  can mine from a little physics, a lot of %%@
philosophy---at least,  a lot more than one paper! To be brief enough, %%@
I shall have to be selective in various ways. The two main ones %%@
are:--\\
\indent 1) I shall consider only a limited class of classical %%@
mechanical systems, and give a technically elementary presentation of %%@
how the Lagrangian, Hamiltonian and Hamilton-Jacobi approaches treat %%@
them (Section \ref{TechPrelpsa}). To be a bit more specific: I shall %%@
consider only systems with finitely many degrees of freedom, for which %%@
any constraints can be solved; and my presentation will eschew modern %%@
geometry. This limitation is largely a matter of brevity and expository %%@
convenience: most of the philosophical discussion in Sections %%@
\ref{Threepsa} et seq. applies more widely.\\
\indent 2) These modal involvements are entangled, technically and %%@
philosophically, with the fact that these three approaches provide %%@
general schemes for solving problems, or for representing their %%@
solutions. I believe these general schemes hold philosophical morals. %%@
But here I will set them aside. (My (2003) takes them up.)
  
The plan of the paper is as follows. In Section \ref{TechPrelpsa}, I %%@
review elements of analytical mechanics. Since philosophers are often %%@
familiar with elementary Lagrangian and Hamiltonian mechanics, but %%@
rarely with Hamilton-Jacobi theory, I will give more detail about the %%@
latter. In Section \ref{Threepsa}, I begin my discussion of modality. I %%@
distinguish three grades of modal involvement, according to which kind %%@
of actual matters of fact they allow to vary counterfactually. The %%@
first grade considers counterfactual initial and/or final conditions, %%@
but keeps fixed the forces on the system and the laws of motion. It is %%@
most strikingly illustrated by Hamilton-Jacobi theory's $S$-function, %%@
which represents a structured ensemble of such conditions. The theory %%@
considers many different $S$-functions, and so ensembles: so I discuss %%@
the structure of this set in Section \ref{hjthypsa}. In particular, %%@
there is an analogy with Lewis' spheres of worlds. The third grade of %%@
modal involvement, which considers counterfactual  laws of motion, is %%@
illustrated by the way variational principles, such as Hamilton's %%@
Principle, invoke evolutions that violate the actual law. This prompts %%@
a discussion (Section \ref{Threatpsa}) whether variational principles %%@
violate the philosophical  principle that any actual truth is made true %%@
by  actual facts. I argue that they do not, at least for simple %%@
mechanical systems. But the topic brings out various points, including %%@
another analogy with Lewis' account of counterfactuals. It also raises %%@
some open questions.

\section{Technical preliminaries}\label{TechPrelpsa}
I will first review the mathematics and physics I need; (without %%@
proofs, but with a few references). Much of what follows is pure %%@
mathematics, though I will use a notation and jargon suggestive of %%@
mechanics. Each of the three approaches---Lagrangian, Hamiltonian and %%@
Hamilton-Jacobi---has a Subsection. 

\subsection{Simple systems and Lagrangian mechanics}
\label{ssec;cvreview}
I begin with the simplest problem of the calculus of variations. This %%@
is the variational problem (in a notation suggestive of mechanics)
\be
\dd I := \dd I[q_i] = \dd \int^{t_1}_{t_0} L(q_1,\dots,q_n,{\dot %%@
q}_1,\dots,{\dot q}_n,t) dt = 0 \;\; , \;\;\; 
\label{actionwithpara=t}
\ee
where $[\;]$ indicates that $I$ is a functional, the dot denotes %%@
differentiation with respect to $t$, and 
$L$ is to be a $C^2$ (twice continuously differentiable) function in %%@
all $2n+1$ arguments. $L$ is the {\em Lagrangian} or {\em fundamental %%@
function}; and $\int L \; dt$ is the {\em action} or {\em fundamental  %%@
integral}. I will discuss this only locally; i.e. I will consider a %%@
fixed simply connected region $G$ of $(n+1)$-dimensional real space %%@
$\mathR^{n+1}$, on which there are coordinates $(q_1,\dots,q_n,t) =: %%@
(q_i,t)$. I will often suppress the subscripts $i,j$ etc. running from %%@
1 to $n$, and write $(q,t)$ etc. 

The singling out of a coordinate $t$ (called the {\em parameter} of the %%@
problem), to give a parametric representation of curves $q(t) := %%@
q_i(t)$, is partly a matter of notational clarity. But it is of course  %%@
suggestive of the application to mechanics, where  $t$ is time, $q$ %%@
represents the system's configuration and $(q_i,t)$-space is often %%@
called `extended configuration  space' or `event space'. Besides, the %%@
singling out of $t$  reflects the fact that we do not require the %%@
fundamental integral to be independent of the choice of $t$; indeed we %%@
shall note in Section \ref{ssec;canleqnsHMpsa} that allowing this %%@
dependence is necessary for making Legendre %%@
transformations.\footnote{Of course, the calculus of variations {\em %%@
can} be developed on the assumption that the fundamental integral is to %%@
be parameter-independent---if it could not be, so much the worse for %%@
relativistic theories! But the details, in particular of how to set up %%@
a canonical formalism, are different from what follows in this Section, %%@
and I set them aside; (cf. e.g. Rund (1966, Chapter 3)). Suffice it to %%@
say that the philosophical morals of Sections 3 et seq. hold good for %%@
parameter-independent problems.}

A necessary condition for $I$ to be stationary at the $C^2$ curve $q(t) %%@
:= q_i(t)$---i.e. for $\dd I = 0$ in comparison with other $C^2$ curves %%@
that  (i) share with $q(t)$ the end-points $q(t_0), q(t_1)$ and (ii) %%@
are close to $q(t)$ in both value and derivative throughout $t_0 < t < %%@
t_1$---is that: $q(t)$ satisfies for $t_0 < t < t_1$ the $n$ %%@
second-order {\em Euler-Lagrange} (also known as: {\em Lagrange}) %%@
equations
\be
\frac{d}{dt}L_{{\dot q}_i} - L_{q_i} = 0 \;\;\;\; i = 1,\dots,n,
\label{elpara=t}
\ee
where as usual subscripts indicate partial differentiation; i.e. %%@
$L_{{\dot q}_i}: = \frac{\pl L}{\pl {\dot q}_i}$ etc. The proof is %%@
elementary. Under certain conditions, the converse also holds: that is, %%@
eq. \ref{elpara=t} is sufficient for eq. \ref{actionwithpara=t}, i.e. %%@
for $I$ to be stationary .  
A curve satisfying eq. \ref{elpara=t} is called an {\em extremal}. 

We apply these ideas to mechanics, getting {\em Lagrangian mechanics}. %%@
We consider a mechanical system with $n$ configurational degrees of %%@
freedom. Note that if the system consists of $N$ point-particles (or %%@
bodies small enough to be treated as point-particles), so that a %%@
configuration is fixed by $3N$ cartesian coordinates, we may yet have %%@
$n < 3N$; for the system may be subject to constraints and the $q_i$ %%@
are to be independently variable in the region $G$.

I shall assume that the system is {\em simple}, in the sense that it %%@
has the following five features, (i) to (v). Note: (1): My discussion %%@
of the Hamiltonian and Hamilton-Jacobi  approaches will retain this %%@
restriction to simple systems; and each will also assume other %%@
restrictions. (2): Some of these features, e.g. (iii), evidently %%@
involve modal notions; but I will postpone discussion of these aspects %%@
till Section \ref{Threepsa} et seq..\\
\indent  (i): Any constraints on the system are {\em holonomic}; i.e. %%@
each is expressible as an equation $f({r}_1,\dots,{r}_m) = 0$ among the %%@
coordinates ${r}_k$ of the system's component parts; (here the $r_k$ %%@
could be the $3N$ cartesian coordinates of $N$ point-particles, so that %%@
$m := 3N$). A set of $c$ such constraints can in principle be solved, %%@
defining a $(m - c)$-dimensional hypersurface $Q$ in the %%@
$m$-dimensional space of the $r$s; so that on the {\em configuration %%@
space} $Q$ we can define $n := m-c$ independent coordinates $q_i, i = %%@
1,\dots,n$.\\
\indent (ii): Any constraints on the system are {\em scleronomic}, i.e. %%@
independent of time. So the configuration space $Q$ is identified once %%@
and for all; and we can take the region $G \subset \mathR^{n+1}$ as a %%@
cartesian product of $Q$ with a time-interval $[t_-,t_+] \subset %%@
\mathR$ (where we allow $t_- = -\infty, t_+ = +\infty$).\\
\indent (iii): Any constraints on the system are {\em ideal}; i.e. the %%@
forces that maintain the constraints would do no work in any possible %%@
displacement consistent with the constraints and applied forces (called %%@
a {\em virtual displacement}). This allows us to deduce the principle %%@
of virtual work, and thereby d'Alembert's principle.\\
\indent D'Alembert's principle implies that for a holonomic system %%@
(i.e. obeying (i)), the {\em kinetic energy} $T$ (defined in cartesian %%@
coordinates, with $k$ now labelling particles, by: $T := \Sigma_k %%@
\frac{1}{2}m_k{\bf v}^2_k$) and {\em generalized forces} $Q_i$ (which %%@
are defined for $i = 1,\dots,n$, in terms of the vector applied force %%@
${\bf F}_k$ on particle $k$, and position vector ${\bf r}_k$ of %%@
particle $k$, by: $Q_i := \Sigma_k {\bf F}_k \cdot \left(\frac{\pl {\bf %%@
r}_k}{\pl q_i}\right)$) obey, for all $i$
\be
\frac{d}{dt}\left(\frac{\pl T}{\pl {\dot q}_i}\right) - \frac{\pl %%@
T}{\pl q_i} \;\; \equiv \;\; 
\frac{d}{dt}\left(T_{{\dot q}_i}\right) - T_{q_i} \;\; = \;\; Q_i %%@
\;\;\; ;
\label{LageqnTQ} 
\ee 
which are also sometimes called Lagrange's equations. \\
\indent  (iv): The applied forces are {\em monogenic}; i.e. the total %%@
work $\dd w$ done in an infinitesimal virtual displacement is %%@
integrable; its integral is the {\em work function} $U$.  (The term %%@
`monogenic' is due to Lanczos (1986, p.30), but followed by others e.g. %%@
Goldstein et al. (2002, p. 34).) \\
\indent (v): Furthermore, the system is {\em conservative}; i.e. the %%@
work function $U$ is independent of both the time and the generalized %%@
velocities ${\dot q}_i$, and depends only on the $q_i$: $U = %%@
U(q_1,\dots,q_n)$. We interpret $V:= -U$ as {\em potential energy}. %%@
Then (ii) and (v) together imply the conservation of energy, i.e. the %%@
constancy in time of $T + V$.

 Besides, (v) and the definition of $Q_i$ in (iii) implies that $Q_i = %%@
- V_{q_i}$; so that, defining the Lagrangian $L := T - V$, eq. %%@
\ref{LageqnTQ} take on the form of the Euler-Lagrange equations, i.e. %%@
eq. \ref{elpara=t}. With this $L \equiv T - V$, eq. \ref{elpara=t} are %%@
called Lagrange's equations.

For a simple system, Lagrange's equations are (not just necessary but %%@
also) {\em sufficient} for the action integral $I = \int L \; dt$ to be %%@
stationary (Whittaker (1959, Section 99)). So we infer {\em Hamilton's %%@
Principle}: that the motion   in configuration space of a simple %%@
system, between prescribed configurations at times $t_0$ and $t_1$, %%@
makes stationary $\int L \; dt$,  with the Lagrangian $L(q_i,{\dot %%@
q}_i,t) \equiv L(q_i,{\dot q}_i) := T - V$ now having no explicit %%@
time-dependence:  
\be
\dd I  = \dd \int^{t_1}_{t_0} L(q_i,{\dot q}_i) dt = 0 \;\; .
\label{hpsimple}
\ee

As I mentioned in Section \ref{Intrpsa}, my restriction to simple %%@
systems is largely a matter of brevity and expository convenience, not %%@
of substance. Most of both the formalism below, and the philosophical %%@
morals of later Sections, apply much more widely. For example, in the %%@
last paragraph's deduction of eq. \ref{elpara=t}, the assumption of %%@
conservativity, (v), could be weakened so as to allow $V$ to have %%@
explicit time-dependence and even some forms of velocity-dependence; %%@
(cf. e.g. Goldstein et al. (2002, p. 22); hence eq. %%@
\ref{actionwithpara=t}'s allowance of $t$ as an argument of $L$.).

But beware: some points in what follows {\em are} restricted. The most %%@
important example concerns the deduction of Hamilton's Principle from %%@
Lagrange's equations eq. \ref{elpara=t}; (cf. the last paragraph but %%@
one). This deduction depends on the system being simple; (more %%@
specifically, on the constraints being holonomic, cf. Papastavridis %%@
(2002, pp. 960-973)). We shall see in Section \ref{Threatpsa} that this %%@
leaves us open questions about the modal involvements of Lagrangian and %%@
Hamiltonian mechanics for non-simple systems.

Finally, I note that the power of Lagrangian mechanics as a scheme for %%@
solving problems arises in large part from its equations being %%@
invariant under arbitrary transformations, with non-vanishing Jacobian, %%@
of the $q_i$ (called {\em point transformations}). Thus we are free to %%@
use coordinates $q_i$ to suit the problem at hand: the equations of %%@
motion will retain the form eq. \ref{elpara=t}. 

\subsection{Canonical equations and Hamiltonian mechanics}
\label{ssec;canleqnsHMpsa}
Under certain conditions, the variational problem eq. %%@
\ref{actionwithpara=t} has an equivalent form, {\em the canonical %%@
form}, for which the Euler-Lagrange equations are $2n$ first order %%@
equations, rather than $n$ second order equations; as follows. Starting %%@
from eq. \ref{actionwithpara=t}, we define new variables
\be
p_i := L_{{\dot q}_i} \; ,
\label{definepsubi}
\ee 
called {\em (canonical) momenta}, since in mechanics examples they %%@
often coincide with momenta. Recalling that $L$ is $C^2$ in all its %%@
arguments, we now assume that the Hessian with respect to the ${\dot %%@
q}$s does not vanish in the domain $G$ considered, i.e. the determinant 
\be
\mid L_{{\dot q}_i{\dot q}_j}\mid \; \neq 0 \;\; ;
\label{nonzerohessian}
\ee
so that eq. \ref{definepsubi} can be solved for the ${\dot q}_i$ as %%@
functions of $q_i, p_i,t: {\dot q}_i = {\dot q}_i(q_j,p_j,t)$. Then the %%@
equations
\be
p_i = L_{{\dot q}_i} \;\;\; {\dot q}_i = H_{p_i} \;\;\;
L(q_i,{\dot q}_i,t) + H(q_i,p_i,t) = \Sigma_i {\dot q}_ip_i
\label{legtrfmnsymmicform2nd}
\ee 
represent a {\em Legendre transformation} and its inverse; where in the %%@
third equation ${\dot q}_i$ are understood as functions of %%@
$(q_j,p_j,t)$ according to the inversion of eq. \ref{definepsubi}. The %%@
function $H(q_i,p_i,t)$ is called the {\em Legendre {\rm (or:} %%@
Hamiltonian{\rm )} function} of the variational problem, and the $q$s %%@
and $p$s are called {\em canonically conjugate}. It follows that $H$ is %%@
$C^2$ in all its arguments, $H_t = -L_t$, and $\mid L_{{\dot q}_i{\dot %%@
q}_j}\mid \; = \; \mid H_{{p}_i{p}_j}\mid^{-1}$. Besides, any function  %%@
$H(q_i,p_i,t)$ that is $C^2$ in all its arguments, and has a %%@
non-vanishing Hessian with respect to the $p$s, $\mid %%@
H_{{p}_i{p}_j}\mid \neq 0$, is the Legendre function of a  $C^2$ %%@
Lagrangian $L$ that is given in terms of $H$ by eq. %%@
\ref{legtrfmnsymmicform2nd}. 

Applying this Legendre transformation, the Euler-Lagrange equations eq. %%@
\ref{elpara=t} go over to the {\em canonical system} of equations (also %%@
known as: {\em Hamilton's equations}) 
\be
{\dot q}_i = H_{p_i} \;\;\; {\dot p}_i = -H_{q_i} \; (=L_{q_i}) \;\;\; %%@
.
\label{canleqnswithinCV2nd}
\ee
(A curve satisfying these equations is also called an extremal.) 

Furthermore, these are the Euler-Lagrange equations of a variational %%@
problem equivalent to the original one, in which both $q$s and $p$s are %%@
varied independently, namely
the problem
\be
\dd \int \left(\Sigma_i {\dot q}_ip_i - H(q_i,p_i,t)\right) \; dt \;\; %%@
= \;\; 0 \;\; .
\label{canlvarnprobwithinCV2nd} 
\ee 
The reason for the equivalence, in brief, is:-- The variation of $L = %%@
\Sigma_i {\dot q}_ip_i - H$ with respect to $p_i$ gives $\dd L = %%@
\Sigma_i({\dot q}_i - \frac{\pl H}{\pl p_i})\dd p_i$. Since the term in %%@
brackets vanishes by Hamilton's equations, an arbitrary variation of %%@
the $p_i$ has no influence on the variation of $L$; so the %%@
Euler-Lagrange equations got by varying the $q$s and $p$s independently %%@
are eq. \ref{canleqnswithinCV2nd}, i.e. the Legendre transform of the %%@
originals, eq. \ref{elpara=t}.\footnote{For more discussion of the %%@
Legendre transformation, cf. e.g.: Arnold (1989, Chap.s 3.14, 9.45.C), %%@
Courant and Hilbert (1953, Chap. IV.9.3; 1962, Chap. I.6), Lanczos %%@
(1986, Chap VI.1-4).) I stress that in the theory of the Legendre %%@
transformation, the assumption of a non-vanishing Hessian, eq. %%@
\ref{nonzerohessian} (equivalently: $\mid H_{{p}_i{p}_j}\mid \neq 0$), %%@
is  crucial; if it fails, we need a different theory (called {\em %%@
constrained dynamics}). Incidentally, it also implies that the %%@
fundamental integral cannot be parameter-independent; cf. e.g. Rund %%@
(1966, pp. 16, 141-144).}

Applying these ideas to the Lagrangian mechanics of a simple system, %%@
understood as in Section \ref{ssec;cvreview}, we get {\em Hamiltonian %%@
mechanics}. Thus we now assume not only that the mechanical system is %%@
simple, but also that eq. \ref{nonzerohessian} holds.  And we think of %%@
the system's state-space as, not $Q$, but the $2n$-dimensional {\em %%@
phase space} $\Gamma$ coordinatized by the $p$s and $q$s; (technically %%@
it is the cotangent bundle of $Q$---but as announced in Section 1, I %%@
eschew modern geometry!). The system's motion is given by the new %%@
variational principle, sometimes called the {\em modified Hamilton's %%@
Principle}, eq. \ref{canlvarnprobwithinCV2nd}; or more explicitly, by %%@
Hamilton's equations, eq. \ref{canleqnswithinCV2nd}. 

The Hamiltonian mechanics of a simple system is equivalent to Section %%@
\ref{ssec;cvreview}'s Lagrangian mechanics, together with eq. %%@
\ref{nonzerohessian}. But it has several advantages over Lagrangian %%@
mechanics, as regards both problem-solving and general theory; though I %%@
only mention two.\footnote{Other advantages of the Hamiltonian %%@
approach, from a physical perspective, include: (a) it can be applied %%@
to systems to which the Lagrangian approach does not apply, i.e. in %%@
modern terms, whose phase space is not a cotangent bundle; (b) it %%@
connects analytical mechanics with other fields of physics, especially %%@
statistical mechanics and optics.}\\
\indent (i): Its use of first-order ordinary differential equations. In %%@
particular,  the initial value problem is straightforward, in that %%@
through a given point %%@
$(q_0,p_0):=(q_{0_1},\dots,q_{0_n};p_{0_1},\dots,p_{0_n}) \in \Gamma$, %%@
there passes a unique solution of eq. \ref{canleqnswithinCV2nd}, i.e. a %%@
unique extremal with $q_i(0)=q_{0_i}, p_i(0)= p_{0_i}$.\\
\indent (ii): Its replacement of the group of point transformations on %%@
$Q$ by what is in effect a larger group of transformations on $\Gamma$, %%@
the canonical transformations. There is a rich and multi-faceted theory %%@
of canonical transformations; (to which there are three main %%@
approaches---generating functions,  symplectic geometry and integral %%@
invariants). But I will not need any details about this.

\subsection{Hamilton-Jacobi theory}
\label{ssec;hjpsa}
I shall discuss Hamilton-Jacobi theory in more detail than Lagrangian %%@
and Hamiltonian mechanics; both because it is less familiar to %%@
philosophers and because we need the detail in order to explore its %%@
modal involvements. In Section \ref{ssec;charhypfields}, I follow in %%@
Hamilton's (1833, 1834) footsteps, introducing the Hamilton-Jacobi %%@
equation via Hamilton's characteristic function (as do many mechanics %%@
textbooks); and then in Section \ref{ssec;gfandci} I discuss %%@
hypersurfaces, congruences and fields. Even so, these details will give %%@
only a limited view of a rich theory. In particular:\\
\indent (1) I will ignore aspects to do with problem-solving %%@
(especially the use of separation of variables, leading on to %%@
action-angle variables and Liouville's theorem) since---though %%@
obviously crucial for physics, and so rightly emphasised in %%@
textbooks---they are not illuminating about modality.\\
\indent (2) I will ignore the integration theory of the Hamilton-Jacobi %%@
equation, which involves the theory of generating functions and %%@
complete integrals. This deep (and beautifully geometric) theory {\em %%@
does} illuminate Hamilton-Jacobi theory's modal involvements; but space %%@
prevents me discussing it here.\\
\indent (3) Both Sections \ref{ssec;charhypfields} and %%@
\ref{ssec;gfandci} will emphasise the description of motion in the %%@
extended configuration space of Section \ref{ssec;cvreview}, i.e. the %%@
region $G \subset \mathR^{n+1}$; while it is equally illuminating to %%@
consider Hamilton-Jacobi theory  in phase space. But this emphasis on %%@
$G \subset \mathR^{n+1}$ will suffice for our purposes---to reveal some %%@
distinctive modal involvements.

\subsubsection{The characteristic function and the Hamilton-Jacobi %%@
equation}
\label{ssec;charhypfields}
We now assume that our region $G \subset \mathR^{n+1}$ is sufficiently %%@
small that between any two ``event'' points $E_1 = (q_{1i},t_1), E_2 = %%@
(q_{2i},t_2)$ there is a unique extremal curve $C$. To avoid double %%@
subscripts, I will in this Section sometimes suppress the $i$, writing %%@
$E_1 = (q_1,t_1), E_2 = (q_2,t_2)$ etc. Then the value of the %%@
fundamental integral along $C$ is a function of the coordinates of the %%@
end-points; which we call the {\em characteristic function} and write %%@
as
\be
S(q_1,t_1; q_2,t_2) = \int^{t_2}_{t_1} L \; dt = 
\int^{t_2}_{t_1} (\Sigma_i p_i{\dot q}_i - H) \; dt =
\int \Sigma_i p_idq_i - Hdt  
\label{definecharfn}
\ee
where the integral is taken along the unique extremal $C$ between the %%@
end-points, and we have used the Legendre transformation  eq. %%@
\ref{legtrfmnsymmicform2nd}.

Making arbitrary small displacements $(\dd q_1,\dd t_1), (\dd q_2,\dd %%@
t_2)$ at $E_1,E_2$ respectively, and using the fact that the integral %%@
is taken along an extremal, we get for the variation $\dd S$ of $S$
\begin{eqnarray}
\dd S := S(q_1 + \dd q_1,t_1 + \dd t_1; q_2 + \dd q_2,t_2 + \dd t_2) - %%@
S(q_1,t_1; q_2,t_2) = \nonumber \\
\frac{\pl S}{\pl t_1}\dd t_1 + \frac{\pl S}{\pl t_2}\dd t_2 + \Sigma_i %%@
\; \frac{\pl S}{\pl q_{1i}}\dd q_{1i} + \Sigma_i \;
\frac{\pl S}{\pl q_{2i}}\dd q_{2i} = 
\left[\Sigma_i \; p_i\dd q_i - H(q_j,p_j,t)\dd t \right]^{t_2}_{t_1} %%@
\;\; .
\label{varninS}
\end{eqnarray}
Since the displacements are independent, we can identify each of the %%@
coefficients on the two sides of the last equation in eq. %%@
\ref{varninS}, getting 
\begin{eqnarray}
\frac{\pl S}{\pl t_2} = -[H(q_i,p_i,t)]_{t=t_2} & , & 
\frac{\pl S}{\pl q_{2i}} = [p_i]_{t = t_2}
\label{identifycoefftslawvaryingaction;final} \\
\frac{\pl S}{\pl t_1} = [H(q_i,p_i,t)]_{t=t_1} & , & 
\frac{\pl S}{\pl q_{1i}} = - [p_i]_{t = t_1}
\label{identifycoefftslawvaryingaction;initial} 
\end{eqnarray} 
in which the $p_i$ refer to the extremal $C$ at $E_1$ and $E_2$.

These equations are remarkable, since they enable us, if we know the %%@
function  $S(q_1,t_1,q_2,t_2)$ to determine all the  motions of the %%@
system that are possible with the given $S$---without solving any %%@
differential equations! For suppose we are given the initial conditions %%@
$(q_1,p_1,t_1)$, i.e. the configuration and canonical momenta at time %%@
$t_1$, and also the function $S$. The $n$ equations $\frac{\pl S}{\pl %%@
q_1} = -p_1$ in eq. \ref{identifycoefftslawvaryingaction;initial} %%@
relate the $n+1$ quantities $(q_2,t_2)$ to the given constants %%@
$q_1,p_1,t_1$. So in principle, we can solve these equations by a %%@
purely algebraic process, to get $q_2$ as a function of $t_2$ and the %%@
constants $q_1,p_1,t_1$; i.e. to get the system's motion. Finally, we %%@
can get the momentum $p_2$ from the $n$ equations $p_2 = \frac{\pl %%@
S}{\pl q_2}$ in eq. \ref{identifycoefftslawvaryingaction;final}. So %%@
indeed the problem is solved without performing integrations, i.e. just %%@
by differentiation and elimination: a very remarkable %%@
technique.\footnote{As Hamilton realized. He writes, in the impersonal %%@
style of the day, that `Mr Hamilton's function $S$ ... must not be %%@
confounded with that so beautifully conceived by Lagrange for the more %%@
simple and elegant expression of the known differential equations [i.e. %%@
$L$]. Lagrange's function {\em states}, Mr Hamilton's function would %%@
{\em solve} the problem. The one serves to form the differential %%@
equations of motion, the other would give their integrals' (1834, p. %%@
514).} 

In fact we can from now on ignore the initial time equations eq. %%@
\ref{identifycoefftslawvaryingaction;initial} and study only the $n+1$ %%@
final time equations   eq. \ref{identifycoefftslawvaryingaction;final}. %%@
Roughly speaking, the reason is that eq. %%@
\ref{identifycoefftslawvaryingaction;final} contains enough information %%@
for us to analyse fully the system's motion.\footnote{This insight is %%@
essentially due to Jacobi. For discussion, cf. Dugas (1988, p. 401), %%@
Lanczos (1986, pp. 225, 231-34, 254-57).} So we will often write $t$ %%@
rather than $t_2$ and $q$ rather than $q_2$.

Substituting the second set of equations of eq.  %%@
\ref{identifycoefftslawvaryingaction;final} in the first, and rewriting %%@
$t_2, q_2$ as $t, q$, yields 
\be
\frac{\pl S}{\pl t} + H(q,\frac{\pl S}{\pl q}, t) = 0 \;\; .
\label{hjfinalfromvaryingaction}
\ee
This first order partial differential equation is the {\em %%@
Hamilton-Jacobi equation}; it is non-linear since the contribution of %%@
the kinetic energy $T$ to the Hamiltonian will contain $p^2$ terms. %%@
Throughout this Subsection and the next, we will focus on this %%@
equation.

By fixing $E_1 = (q_1,t_1)$ and considering different values for $S$, %%@
we also see that $S$ defines a family of hypersurfaces, which we can %%@
call `spheres' with centre $E_1 = (q_1,t_1)$. Thus the sphere with %%@
radius $R$ is given by the equation
\be
S(q_1,t_1;q_2,t_2) = R
\label{geodsphereR}
\ee
with $(q_1,t_1)$ considered fixed. Every point $E_2 = (q_2,t_2)$ on %%@
this sphere is connected to the centre $E_1 = (q_1,t_1)$ by a unique %%@
extremal along which the fundamental integral has value $R$. This is %%@
amusingly reminiscent of Lewis' spheres of worlds (1973, Chapter 1.3; %%@
1986, Chapter 1.3): and more than amusingly---we will see in Section %%@
\ref{hjthypsa} that the analogy is deeper.
   
\subsubsection{Hypersurfaces, congruences and %%@
fields}\label{ssec;gfandci}
Of course, partial differential equations have many solutions: (the %%@
main contrast with ordinary differential equations being that %%@
typically, the solution contains an arbitrary function (or functions) %%@
rather than an arbitrary constant (or constants)). So Hamilton-Jacobi %%@
theory studies the whole space of solutions of the Hamilton-Jacobi %%@
equation. I need to report the main classical result of this study. %%@
(For details, a good reference is Rund (1966, Chap. 2), who cites %%@
various masters of the last two centuries, especially Carath\'{e}odory. %%@
I also give more details in (2003a).)

The result connects three diverse notions:---\\
\indent (a): {\em Families of hypersurfaces} in our region $G$ of %%@
$\mathR^{n+1}$
\be
S(q_i,t) = \sigma
\label{famyhyperintrodcd}
\ee
with $\sigma \in \mathR$ the parameter labelling the family; where we %%@
assume that $S$ is a $C^2$ function in all $n+1$ arguments, and that  %%@
the family foliates the region $G$ {\em simply} in the sense that %%@
through each point of $G$ there passes a unique hypersurface in the %%@
family.\\ 
\indent  (b): {\em Congruences of curves} that: (i) cross the %%@
hypersurfaces and fill $G$ simply in the corresponding sense that %%@
through each point of $G$ there passes a unique curve in the %%@
congruence; and (ii) may be parametrically represented by $n$ equations %%@
giving $q_i$ as $C^2$ functions of $n$ parameters $u_{\al}$ and $t$
\be
q_i = q_i(u_{\alpha},t) \;\; , \; \; i = 1,\dots,n \;\; ;
\label{qfromcongruence}
\ee
where each set of $n$ $u_{\alpha} = (u_1,\dots,u_n)$ labels a unique %%@
curve in the congruence.  Thus there is a one-to-one correspondence %%@
$(q_i,t) \leftrightarrow (u_{\alpha},t)$ in appropriate domains of the %%@
variables, with non-vanishing Jacobian
\be
\mid \frac{\pl q_i}{\pl u_{\alpha}} \mid \;\;\ \neq \;\;\ 0 \;\; .
\label{nonzeroJacobqalpha}
\ee
Such a congruence determines tangent vectors $({\dot q}_i,1)$ at each %%@
$(q_i,t)$; and thereby also values of the Lagrangian %%@
$L(q_i(u_{\alpha},t), {\dot q}_i(u_{\alpha},t),t)$ and of the momentum 
\be
p_i = p_i(u_{\alpha},t) = \frac{\pl L}{\pl {\dot q}_i} \;\; .
\label{pfromcongruence}
\ee
\indent (c): {\em Fields}, defined to be a set of $2n \; C^2$ functions %%@
$q_i, p_i$ of $(u_{\alpha},t)$ as in eqs \ref{qfromcongruence} and %%@
\ref{pfromcongruence}, i.e. with the $q$s and $p$s related by $p_i = %%@
\frac{\pl L}{\pl {\dot q}_i}$. So a congruence determines a field, and  %%@
a field determines (by a Legendre transformation, using eq. %%@
\ref{nonzerohessian}) a set of tangent vectors, and so a congruence.\\
\indent Some jargon: (i) If all the curves  of the congruence %%@
determined by a field are extremals, the field is called a {\em field %%@
of extremals}. (ii) We say a field (or its congruence) {\em belongs to} %%@
a family of hypersurfaces given by eq. \ref{famyhyperintrodcd}  iff %%@
throughout the region $G$ the $p_i = \frac{\pl L}{\pl {\dot q}_i}$ of %%@
the field obey the last $n$ equations of eq. %%@
\ref{identifycoefftslawvaryingaction;final}, i.e. iff we have
\be
p_i = \frac{\pl }{\pl q_i}S(q_i,t) = \frac{\pl}{\pl %%@
q_i}S(q_i(u_{\alpha},t),t) \;\; .
\footnote{One can show that a field belongs to a family of %%@
hypersurfaces iff for all indices $\alpha,\beta = 1,\dots,n$, the %%@
Lagrange brackets of the parameters of the field, i.e.
\be
[u_{\alpha},u_{\beta}] := \Sigma_i \left(\frac{\pl q_i}{\pl %%@
u_{\alpha}}\frac{\pl p_i}{\pl u_{\beta}} -
\frac{\pl q_i}{\pl u_{\beta}}\frac{\pl p_i}{\pl u_{\alpha}} \right)
\label{definelagebrackets}
\ee
vanish identically. Warning: the role of Lagrange brackets in this %%@
theory is sometimes omitted even in excellent accounts, e.g. Courant %%@
and Hilbert (1962, Chap. II.9.4).}
\label{p=dSbydq;definebelonging}
\ee
(iii) Finally, we say that a field $q_i = q_i(u_{\alpha},t), p_i = %%@
p_i(u_{\alpha},t)$ is {\em canonical} if the $q_i, p_i$ satisfy %%@
Hamilton's equations eq. \ref{canleqnswithinCV2nd}: equivalently, if %%@
the curves of the congruence determined by the field are extremals. 

So much for definitions; now the result. The following three conditions %%@
on a $C^2$ function $S:G \rightarrow \mathR$ are equivalent:\\
\indent (1): $S$ is a $C^2$ solution (throughout $G$) of the %%@
Hamilton-Jacobi equation
\be
\frac{\pl S}{\pl t} + H(q,\frac{\pl S}{\pl q}, t) = 0 \;\; .
\label{hjnoindices}
\ee 
\indent (2): The field belonging to the $C^2$ function $S:G \rightarrow %%@
\mathR$, i.e. the field defined at each point in $G$ by $p_i = %%@
\frac{\pl S}{\pl q_i}$, is canonical. Equivalently: the curves of the %%@
congruence belonging to $S$ (the congruence defined from the field by %%@
the Legendre transformation) are extremals.\\
\indent (3): The value of the fundamental  integral $\int L \; dt$ %%@
along the curve $C$ of the congruence belonging to $S$, from any point %%@
$P_1$ on the surface $S(q_i,t) = \sigma_1$ to that point $P_2$ on the %%@
surface $S(q_i,t) = \sigma_2$ that lies on $C$, is the {\em same} for %%@
whatever point $P_1$ we choose; and the value is just $\sigma_2 - %%@
\sigma_1$. That is:
\be
\int^{P_2}_{P_1} L \; dt = \sigma_2 - \sigma_1 \; .\footnote{More %%@
precisely: (1) and (3) are exactly equivalent, (1) implies (2); and (2) %%@
implies that $\frac{\pl S }{\pl t} + H(q,\frac{\pl S}{\pl q},t)$ is a %%@
function of $t$ only, and this function can be absorbed into $H$. For %%@
proofs, cf. e.g. Rund (1966, Chap. 2.2-3), Courant and Hilbert (1962, %%@
Chapter II.9.2-4).}
\label{samevalueforallexls}
\ee

In the light of eq. \ref{samevalueforallexls}, we call a family of %%@
hypersurfaces $S = \sigma$ satisfying any, and so all, of these three %%@
conditions {\em geodesically {\rm (or:} geodetically{\rm)} equidistant} %%@
(with respect to the Lagrangian $L$). So the concentric spheres centred %%@
on $E_1 = (q_1,t_1)$ introduced above (eq. \ref{geodsphereR}) are an %%@
example of a geodesically equidistant family. (In fact they are a %%@
``fundamental'' example, in that other families can be ``built'' from %%@
them in ways studied under the names of `Green's functions', and %%@
`Huygens' principle'.) 

To sum up: Any solution $S$ of the Hamilton-Jacobi equation that is %%@
smooth and local (specifically: $C^2$ and defined in a simply connected %%@
region $G$) has level surfaces $S =$ constant which are traversed by a %%@
congruence of extremals so as to make the level surfaces a geodesically %%@
equidistant family.

But so far it is an open question which $n$-dimensional surfaces $M$ in %%@
$G$ are level surfaces of some smooth, say $C^2$, solution $S$ (in %%@
$G$). In fact it can be shown, subject to some mild conditions about %%@
non-vanishing determinants etc., that:\\
\indent (1): any $n$-dimensional surface $M$ is a level surface of a %%@
solution, and this solution is uniquely defined throughout $G$ by its %%@
value on $M$ (say $S = 0$ on $M$); and\\
\indent (2): for any such surface $M$ and any suitably smooth function %%@
$S:M \rightarrow \mathR$, there is a uniquely defined solution on all %%@
of $G$ which restricts on $M$ to the given $S$. (So (2) generalizes (1) %%@
by $M$ not having to be a level surface.)\\
\indent In the jargon: the initial value problem for the %%@
Hamilton-Jacobi equation has locally a solution, that is unique given %%@
suitably smooth prescribed values of $S$. But I shall not go into %%@
details about this. It suffices to state the intuitive idea for the %%@
case where $M$ is a level surface: the solution is ``grown'' from the %%@
given surface by erecting a congruence of curves, transverse to the %%@
surface, and passing along them to mark off a given value of the %%@
fundamental integral $\int L \; dt$; by varying the value, one defines %%@
a geodesically equidistant family and so a solution $S$. (For details, %%@
cf. Rund (1966, Chap. 2.7-8), Benton (1977, Chap. 1) or my (2003a, %%@
Sections 5,6); or more heuristically, Courant and Hilbert (1962, Chap. %%@
II.9.2-5) and Born and Wolf (1999, Appendix I.2-4).)     

Returning finally to mechanics: it is clear that each solution $S$ of %%@
the Hamilton-Jacobi  equation represents a kind of {\em ensemble}, i.e. %%@
a fictitious population of systems (maybe including the actual system). %%@
Thus each solution $S$ represents an ensemble with the feature that at %%@
all times $t$, there is a strict configuration---momentum, i.e. $q-p$, %%@
correlation given by the gradient of $S$. That is, $S$ prescribes for %%@
any given $(q,t)$, a unique $(p,t):= (\frac{\pl S}{\pl q},t)$. 

So much by way of expounding Hamilton-Jacobi  theory. We will return, %%@
after Section \ref{Threepsa}'s introduction of modality, to discuss the %%@
structure of this set of ensembles (set of $S$-functions)---and so the %%@
modal involvements of Hamilton-Jacobi theory. Finally, I emphasise a %%@
point mentioned in Section \ref{Intrpsa}: that this Section's %%@
restrictive assumptions about the systems to be considered (that the %%@
systems be simple, that eq. \ref{nonzerohessian} hold, that any two %%@
points in $G$ be connected by a unique extremal etc.) are largely a %%@
matter of brevity and expository convenience, not of substance. Much of %%@
the formalism above, and the philosophical morals below, apply more %%@
widely.

\section{Grades of modal involvement}\label{Threepsa}
I turn to discussing the modal involvements of our three approaches to %%@
analytical mechanics: Lagrangian, Hamiltonian and Hamilton-Jacobi. In %%@
this Section, I begin with the obvious fact that postulating a space of %%@
possible states (a state-space: $Q$ or $\Gamma$ or $G$ in Section %%@
\ref{TechPrelpsa}'s notation) brings in modality. This leads to a %%@
suggested distinction between three grades of modal involvement, which %%@
are all illustrated by analytical mechanics. In this Section, I just %%@
briefly mention some illustrations: the next two Sections discuss some %%@
more striking illustrations of the first and third grades. 

At first sight, the philosophical import of postulating a state-space %%@
would seem to be at most some uncontroversial version of the idea that %%@
laws support counterfactuals. That is: whether or not one believes in a %%@
firm distinction between laws of nature and accidental generalizations, %%@
and whatever one's preferred account of counterfactuals, a theory that %%@
states `All As are Bs' surely in some sense warrants counterfactuals %%@
like `If any object were an A, it would be a B'. And so when analytical %%@
mechanics postulates state-space and then specifies e.g. laws of motion %%@
on it, it seems at first that this just corresponds to the passage from %%@
`All actual systems of this kind (having such-and-such initial %%@
states---usually a ``small'' proper subset of state-space) evolve %%@
thus-and-so' to `If any system of this kind were in any of its possible %%@
initial states, it would evolve thus-and-so'.\footnote{Here, and in all %%@
that follows, I of course set aside the (apparent!) fact that the %%@
actual world is quantum, not classical; so that I can talk about e.g. %%@
an actual system obeying Hamilton's Principle. Since my business %%@
throughout is the philosophy of classical mechanics, it is unnecessary %%@
to encumber my argument with antecedents like `If the world were not %%@
quantum': I leave you to take them in your stride.}

But this first impression is deceptive. The structures with which %%@
state-space is equipped by analytical mechanics, and the constructions %%@
in which it is involved, make for more varied and nuanced involvements %%@
with modality than is suggested by just the idea that laws support %%@
counterfactuals. In the light of Section \ref{TechPrelpsa}, I think it %%@
is natural to distinguish (in Quinean fashion!) three grades of modal %%@
involvement; so I shall write (Modality;1st) etc. Like Quine's three %%@
grades, the first is intuitively the mildest grade, and the third the %%@
strongest. But this order will not correspond to Section %%@
\ref{TechPrelpsa}'s (and the historical) order of the three approaches %%@
to analytical mechanics. In particular, the first and arguably most %%@
intuitive approach, Lagrangian mechanics, exhibits the {\em third} %%@
grade of modal involvement.\footnote{Also, my three grades (like %%@
Quine's) are often combined: for example, a mechanical theory (or even %%@
a small part of one, such as a theorem) might involve the first and %%@
third grades. But I will not need to discuss such combinations.}

The grades are defined in terms of the kind of actual matters of fact %%@
they allow to vary counterfactually. The first kind is, roughly, the  %%@
state of the system. The second kind is the  physical problem: which we %%@
can take as specified by the number of degrees of freedom, and the %%@
Lagrangian or Hamiltonian which encodes the forces on the system. The %%@
third kind is the laws of motion, as specified by e.g. Hamilton's %%@
Principle or Lagrange's or Hamilton's equations. Thus we have the %%@
following grades.\footnote{I don't claim that these three grades are %%@
the best way to classify the modal involvements of analytical %%@
mechanics. For example, it might be at least as fruitful to consider %%@
how the various kinds of constraint (holonomic, scleronomic etc. and %%@
their contraries) classify the notion of virtual displacement: this %%@
classification would cut right across the trichotomy that follows. But %%@
at least what follows has the merits of being: obvious, suggested by %%@
Section \ref{TechPrelpsa}'s review, and showing at least some of the %%@
variety of modal involvement that occurs.}

(Modality;1st): The first and mildest grade keeps fixed the actual   %%@
physical problem and laws of motion. But it considers different initial %%@
or final conditions than the actual given ones. And so it also %%@
considers  counterfactual histories of the system; (since under %%@
determinism, a different initial or final state implies a different %%@
history, i.e. trajectory in state-space). 

So this grade includes the familiar idea above, that laws support %%@
counterfactuals. But analytical mechanics  also provides more %%@
distinctive illustrations. Some arise from the postulation of a %%@
state-space, be it $Q$ or $\Gamma$ or $G$. Thus recall from Section %%@
\ref{ssec;cvreview}'s definition of a simple system: (a) the %%@
configuration space $Q$ is to have independently variable coordinates %%@
$q_i$; and (b) to define ideal constraints, one needs to define a %%@
virtual displacement (as one that the system could undergo compatibly %%@
with the constraints and applied forces). But the most striking %%@
illustration of (Modality;1st) is Hamilton-Jacobi theory. As we saw in %%@
Section \ref{ssec;hjpsa}, Hamilton-Jacobi theory enables you to  solve %%@
a problem, as it might be an actual one, by introducing an ensemble of %%@
systems; i.e. a set of possible systems, of which the actual system is %%@
just one member. Besides, the ensemble can be chosen in such a way that %%@
given $S$ the problem is solved without performing integrations, i.e. %%@
just by differentiation and elimination. For more discussion, cf. %%@
Section \ref{hjthypsa}.

(Modality;2nd): The second grade keeps fixed the laws of motion, but %%@
considers different problems than the actual one (and thereby different %%@
initial states). For example, it considers a counterfactual number of %%@
degrees of freedom, or a counterfactual potential function. Maybe no %%@
actual system is a simple system with 5,217 coordinates (nor even is %%@
well modelled as one); or with a potential given (in certain units) by %%@
the polynomial $13x^7 + 5x^3 + 42$.
But  analytical mechanics continually considers such counterfactual %%@
cases: in Section \ref{TechPrelpsa}, we generalized from the outset %%@
about the number $n$ of degrees of freedom, and about what the %%@
Lagrangian or Hamiltonian was (subject of course  to  conditions like %%@
eq. \ref{nonzerohessian}). Such generality of course  pays off in %%@
countless general theorems. 

(Modality;3rd): The third grade considers different laws of motion, %%@
even for a given problem. Again, this can happen  even in Lagrangian %%@
mechanics; namely in its use of Hamilton's Principle eq. %%@
\ref{actionwithpara=t}, and eq. \ref{hpsimple} for simple systems. It %%@
also happens in Hamiltonian mechanics with its modified Hamilton's %%@
Principle, eq. \ref{canlvarnprobwithinCV2nd}. (And Section %%@
\ref{ssec;hjpsa} showed that these variational principles are  also %%@
involved in the Hamilton-Jacobi approach.) In all three approaches, the %%@
use of variational principles means---not that one explicitly states %%@
non-actual laws, much less calculates with them---but that one states %%@
the actual law as a condition that compares the actual history of the %%@
system  with counterfactual histories of it that do not obey the law %%@
(in philosophers' jargon: are {\em contralegal}). That is, the %%@
counterfactual histories share the initial and final conditions, but do %%@
not obey the given deterministic laws of motion, with the given %%@
forces.\footnote{Besides, one does not require that there {\em could} %%@
be forces which in conjunction with the actual laws and initial and %%@
final conditions, would yield the counterfactual history. But I agree %%@
that in general for each suitably smooth counterfactual history, there %%@
will be some recipe of (in general time-dependent) internal and %%@
external forces that would yield the history: (thanks to Michael %%@
Dickson for pointing this out). So I also agree that my distinction %%@
between (Modality;3rd) and (Modality;2nd), between varying the laws and %%@
varying the forces, is not as hard-and-fast as it first seems: one can %%@
in this sense avoid (Modality;3rd) by accepting (Modality;2nd). But %%@
this point will not affect my discussion.} This is at first sight %%@
surprising, even mysterious. How can it be possible to state the actual %%@
law by a comparison of the actual history with possible histories that %%@
do not obey it? I take up this question in Section \ref{Threatpsa}.

To sum up, analytical mechanics gives many illustrations of all three %%@
grades: indeed, the subject is upto its ears in modality. But rather %%@
than multiplying examples, the remainder of this paper undertakes two %%@
projects suggested by my trichotomy. The first (Section \ref{hjthypsa}) %%@
concerns Hamilton-Jacobi theory's use of (Modality;1st). There is no %%@
special philosophical {\em difficulty} here; rather the situation %%@
represents an invitation to philosophers  to study a new sort of modal %%@
structure. The second project (Section \ref{Threatpsa}) concerns %%@
variational principles, especially in Lagrangian and Hamiltonian %%@
mechanics. Here there is a philosophical difficulty: the variational %%@
principles threaten a plausible philosophical principle, and the threat %%@
needs to be answered.  It can be answered, at least for simple systems; %%@
but doing so pays dividends. 

\section{On the set of ensembles}\label{hjthypsa}
Since the $S$-function, representing an ensemble of systems whose $q$ %%@
and $p$ are correlated by $p = \frac{\pl S}{\pl q}$, stands at the %%@
centre of Hamilton-Jacobi theory, it is clear that the theory  %%@
illustrates (Modality;1st) in spades. As discussed at the end of %%@
Section \ref{ssec;hjpsa}, the structure of the set of ensembles is %%@
essentially given by the structure of the set of suitably smooth (say %%@
$C^2$) real functions on a $n$-dimensional manifold $M$; ($M$ needs to %%@
``lie across'' the region $G$ so as to be transverse to a congruence of %%@
extremals). For since there is a locally unique solution to the %%@
Hamilton-Jacobi initial value problem, each such function %%@
determines---as well as is determined by---a solution  throughout $G$ %%@
of the Hamilton-Jacobi equation. So one infers that the set of %%@
solutions (ensembles) is some kind of infinite-dimensional set. 

This set has various kinds of structure, and a full discussion would %%@
have to take account of the aspects listed at the start of Section %%@
\ref{ssec;hjpsa}, that I am setting aside.  In particular, %%@
Hamilton-Jacobi integration theory (especially the notions of complete %%@
integral, and Jacobi's theorem) picks out subsets of the solution space %%@
which are significant, both theoretically and for problem-solving. But %%@
even with just the results of Section \ref{ssec;hjpsa}, we can discern %%@
two kinds of structure---which bear on Lewis' account of modality, %%@
especially counterfactuals. These two kinds of structure arise from two %%@
different choices about what to take as the analogue, in %%@
Hamilton-Jacobi  theory, of a Lewisian possible world.  

\subsection{Configurations as worlds}\label{configworldpsa}
Let us think of an event (i.e. instantaneous configuration) $(q_i,t) %%@
\in G$ as like a possible world. Then Hamilton's characteristic %%@
function eq. \ref{definecharfn}, and the geodesic spheres it defines %%@
eq. \ref{geodsphereR},  yield a neat analogy with Lewis' theory of %%@
counterfactuals. 

For recall Lewis' proposed truth-conditions for a counterfactual `If %%@
$A$ were the case, then $C$ would be the case', which I will write as %%@
$A \rightarrow C$ (1973, Chap. 1.3). Roughly speaking, he proposes that %%@
$A \rightarrow C$ is true at the actual world @ iff: the possible world %%@
(or worlds) most similar (for short: ``closest'') to @ that makes $A$ %%@
true, also makes $C$ true. But he wants to avoid the  assumption that %%@
there is a set of $A$-worlds tied for first equal as regards similarity %%@
to @ (the Limit Assumption). He also allows the counterfactual to be %%@
vacuously true: namely iff no world in the union of nested spheres %%@
around @, $\cup \; \$_{@}$, makes $A$ true. That is, Lewis proposes %%@
that the counterfactual $A \rightarrow C$ is true at @ iff:--\\
\indent 1) no $A$-world belongs to any sphere $S$ in the system %%@
$\$_{@}$ of spheres around @;\\
\indent or\\
\indent 2) some sphere $S$ in the system $\$_{@}$ contains at least one %%@
$A$-world, and $A \supset C$ is true at every world in $S$; (i.e. $C$ %%@
is true at every $A$-world in $S$).

We can easily transplant this kind of  truth-condition to geodesic %%@
spheres; i.e. taking points $(q_i,t) \in G$ as worlds and $\int L \; %%@
dt$ as the measure of distance (dissimilarity) between such worlds. %%@
However, the resulting conditionals arguably do not deserve the name %%@
`counterfactual', since both the ``base-world'' $(q_1,t_1)$ and the %%@
``closest $A$-world'', say $(q,t)$, that the evaluation of the %%@
conditional carries us to, could be actual configurations of the %%@
system.

For simplicity I will ignore the vacuous case, 1) above. This yields %%@
the following truth-condition, relative to a given configuration %%@
$(q_1,t_1)$:
\begin{quote}       
a) $A$ is true at a possible configuration $(q,t)$, to which the given %%@
configuration $(q_1,t_1)$ could evolve (i.e. would evolve for some %%@
$p_1$ at $t_1$) with $t > t_1$; and\\
b) for every possible configuration $(q',t')$ to which $(q_1,t_1)$ %%@
could evolve with $t > t_1$, and such that $\int^{q',t'}_{q_1,t_1} L \; %%@
dt \leq \int^{q,t}_{q_1,t_1} L \; dt$:\\
$A \supset C$ is true at $(q',t')$; (i.e. if $A$ is true at $(q',t')$, %%@
so is $C$).
\end{quote}

In the abstract, this truth-condition seems a mouthful. But in fact %%@
mechanics provides countless examples of such conditional propositions, %%@
though of course in a much less formal guise! A very simple example is %%@
given by a bead sliding on a wire that lies in a vertical plane; (to be %%@
a simple system in Section \ref{TechPrelpsa}'s sense, the bead must %%@
slide frictionlessly). We can take $A$ to say that the bead is at the %%@
lowest point of the  wire, and $C$ to say that its potential energy is %%@
at a minumum. Then $A \rightarrow C$ can be expressed informally as %%@
`Whenever the bead is next at the lowest point of the  wire, its %%@
potential energy will then be at a minimum'. Similarly, with $C$ saying %%@
instead that the kinetic energy is at a maximum; and so on.

Finally, the results in Section \ref{ssec;gfandci} (especially %%@
condition (3)) implies that this discussion of counterfactuals can be %%@
generalized so as to define similarity of worlds in terms of level %%@
surfaces of any solution $S$ of the Hamilton-Jacobi equation. For %%@
example, we could take a $n$-dimensional surface $M$ that is %%@
topologically a sphere surrounding some given point $(q_1,t_1) \in G$, %%@
define $M$ to be a surface of constant $S$, say $S = 0$, and consider %%@
the (locally unique) solution of the Hamilton-Jacobi equation thereby %%@
defined outside $M$. That is, we could define the dissimilarity of our %%@
worlds $(q,t)$ from the base-world $(q_1,t_1)$, and so the %%@
truth-conditions of counterfactuals, in terms of the value of $S(q,t)$.

\subsection{States as worlds}\label{stateworldpsa}
 On the other hand, let us take as the analogue of a Lewisian world an %%@
instantaneous state in the sense of a $2n+1$-tuple $(q_i,p_i,t)$. This %%@
is perhaps a more natural choice than Section \ref{configworldpsa}'s %%@
instantaneous configurations (events),  since it determines a history, %%@
i.e. a phase space trajectory, of the system, our ``toy-universe''. %%@
(Indeed, an even closer analogue to a Lewisian world would be such a %%@
trajectory, which is equivalent to a one-parameter  family of tuples %%@
$(q_i,p_i,t)$ parameterized by time; but I will not separately discuss %%@
this.) 

As in Section \ref{configworldpsa}, there are various constructions one %%@
could  make with this concept of world. In particular, one could define %%@
conditionals $A \rightarrow C$ by using any solution  $S$ of the %%@
Hamilton-Jacobi equation to define dissimilarity. These conditionals %%@
would in general be counterfactual, since the ``base-world'' %%@
$(q_1,p_1,t_1)$ will be on a different phase space trajectory than the %%@
$(q_2,p_2,t_2)$ that evaluation of the conditional carries us to. But I %%@
shall not pursue this; (partly for the sake of variety---for I will %%@
anyway return to counterfactuals in Section \ref{philosophicalpsa}). I %%@
shall instead describe how an $S$-function enables us to define various  %%@
sets of possible worlds which are ``preferred'' relative to our choice %%@
of $S$; in fact, the last of these definitions is important for %%@
physics.

Here again, the $S$-function can be any solution of the Hamilton-Jacobi  %%@
equation. Given such an $S$, every point $(q,t) \in G$ has an %%@
associated canonical momentum, viz. $p := \frac{\pl}{\pl q}S(q,t)$, and %%@
so an associated world in our sense, viz. $(q,p \equiv \frac{\pl S}{\pl %%@
q},t)$. If we wish, we can also pick out subsets so that not every %%@
event $(q,t)$ is included in a world ``preferred'' by our $S$. For %%@
example, we could do this by picking out a sub-manifold $M$ of $G$, and %%@
defining the associated worlds $(q,p \equiv \frac{\pl S}{\pl q},t)$ %%@
only for $(q,t) \in M$.

There are two obvious ways to specify such an $M$; both make $M$ %%@
$n$-dimensional. First, we can define $M$ as the level surface of $S$ %%@
that passes through some given $(q,t) \in G$. This definition will %%@
connect $M$ with Section \ref{ssec;gfandci}'s discussion of %%@
geodesically equidistant hypersurfaces. And thinking of $(q,t)$ as the %%@
system's actual present configuration, $M$ defines a preferred set of %%@
counterfactual events, i.e. instantaneous configurations (which are in %%@
general not simultaneous with $(q,t)$).

Secondly, we can fix $t$. (This will mean that our worlds are given in %%@
effect by just $(q,p)$ not $(q,p,t)$.) Each value of $t$ defines $M$ as %%@
$Q \times \{t\}$, i.e. the copy of the configuration space $Q$ at time %%@
$t$; (cf. Section \ref{ssec;cvreview}'s assumption (ii), of scleronomic %%@
constraints). Now writing this copy simply as $Q$, we consider the %%@
gradient $\frac{\pl }{\pl q}S(q,t)$ as a function on $Q$.  The  %%@
preferred worlds are then given by all $(q,p(q) \equiv \frac{\pl }{\pl %%@
q}S(q,t))$ for $q \in Q$. So the worlds are given as before, except %%@
that the fixed value of $t$ is now implicit in the definition  of $p$. 

In fact this second definition is crucially important for the %%@
mathematics and physics of Hamilton-Jacobi theory in phase space. For %%@
consider the graph of the function $q \mapsto p(q) := \frac{\pl }{\pl %%@
q}S(q,t)$ in the usual logician's sense of the set of ordered pairs of %%@
arguments and values; that is, consider the set of pairs $(q,p(q))$. It %%@
is a $n$-dimensional surface in the $2n$-dimensional phase space %%@
$\Gamma$. It turns out that it is an example of a special kind of %%@
surface, called Lagrangian submanifolds. I shall not define this %%@
notion: here it suffices to remark that it is crucial for %%@
understanding:\\
\indent (i) the general (symplectic) structure of Hamiltonian mechanics %%@
and Hamilton-Jacobi theory;\\
\indent (ii) physical phenomena like focussing and caustics; these %%@
arise when the assumption we made at the start of Section %%@
\ref{ssec;charhypfields}, that any two events $(q_1,t_1), (q_2,t_2) \in %%@
G$ are connected by a unique extremal, breaks down;\\
\indent (iii) the relation of classical and quantum mechanics, since a %%@
Lagrangian submanifold is in effect the  classical analogue of a pure %%@
quantum state (taken as an assignment of values to a complete set of %%@
observables).\footnote{For more details about Lagrangian submanifolds, %%@
cf. e.g. Arnold (1989, Chap.s 7,8), Littlejohn (1992, Sections 1-3).}\\
For us, the main point is, as before, about the structure of the modal %%@
involvement. Namely: the graph of $q \mapsto p(q)$ gives us a preferred %%@
set of worlds, i.e. alternative states in phase space. Besides, we can %%@
analyse the structure of the set of possible preferred sets by studying %%@
the set of all Lagrangian manifolds; (or instead, its quotient by the %%@
time-evolution under the Hamiltonian $H$).   
 
So much by way of surveying the structure of Hamilton-Jacobi theory's %%@
set of ensembles---surveying the riches of (Modality;1st). I close this %%@
Section with a philosophical remark, which looks ahead to Section %%@
\ref{Threatpsa}. There I will deny that merely possible facts (or %%@
states of affairs or other ``truthmakers'') could be what make true an %%@
actually true proposition; (for, I will claim, only actual facts could %%@
do that). But for all I have so far said about Hamilton-Jacobi theory, %%@
one might think that it involves precisely this idea---which Lewis once %%@
jokingly called ``possibilia power'' (1986a, p. 158). After all, what %%@
else might the use of an $S$-function i.e. an ensemble of possible %%@
systems (for example, to solve a problem) come to?

But in fact, there is no conflict. Agreed, Hamilton and Jacobi teach us %%@
to use an $S$-function to solve problems; and for a single problem %%@
there are many $S$-functions we can consider (which do not all differ %%@
just by the time-parameter). But there is no strange influence (whether %%@
causal or constitutive) of the $S$-function, or the ensemble it %%@
represents, on the actual system (or propositions about it). In %%@
particular, the evolution of a system (its trajectory in configuration %%@
space or phase space) {\em is} fixed by, for example, the initial %%@
conditions---$q,{\dot q},t$ in Lagrangian mechanics and $q,p,t$ in %%@
Hamiltonian mechanics---irrespective of which if any $S$-function  we %%@
care to use.\footnote{Incidentally, the situation is different in %%@
quantum theory. There, $S$ has a close mathematical cousin (also %%@
written $S$) whose values do influence the motion of the system. But %%@
again, this does not represent any weird ``possibilia power''. For this %%@
influence is regarded as a strong, indeed the strongest, reason to take %%@
the quantum $S$-function as part of the actual physical state of the %%@
individual system; i.e. not as in classical mechanics, as ``just'' a %%@
description of an ensemble.}

\section{Truths without truthmakers?}\label{Threatpsa}
As I said at the end of Section \ref{Threepsa}, it seems odd, even %%@
mysterious, to state an actual dynamical law by a comparison of the %%@
actual history with possible histories that do not obey it---yet %%@
variational principles do just this. I shall argue that in fact there %%@
is no problem here. But the topic repays scrutiny: it yields insights, %%@
both philosophical (Section \ref{philosophicalpsa}) and technical %%@
(Section \ref{cvrevisitedpsa}); and it raises some open %%@
questions.\footnote{The topic seems wholly ignored in the philosophical %%@
literature about variational principles. But thanks to the rise of %%@
modal metaphysics in analytical philosophy---over which Lewis presided %%@
so magnificently---the topic is nowadays plainly visible. Incidentally: %%@
the literature has instead focussed almost entirely on the way (i) %%@
specifying final conditions and (ii) referring to {\em least} action, %%@
suggests teleology. Indeed, this focus has been dominant ever since %%@
Maupertuis (cf. e.g. Yourgrau and Mandelstam (1979, Chap. 14), Dugas %%@
(1988)). But I set it aside entirely.}

I shall concentrate on Lagrangian mechanics, and specifically on %%@
Hamilton's Principle. Recall that for the simple systems we are %%@
concerned with, this states that the motion between prescribed %%@
configurations
at time $t_0$ and time $t_1$ makes stationary the action integral:
\be
\dd I = \dd \int^{t_1}_{t_0} L(q_1,\dots,q_n,{\dot q_1},\dots,{\dot %%@
q_n}) \; dt = 0 \; .
\ee 
This involves (Modality;3rd): not because it formulates non-actual %%@
laws, but because of the kind of variation it uses to state the actual %%@
law.

I say `shall concentrate' for two reasons, the first ``positive'' and %%@
the second ``negative''. (1): I shall also mention the modified %%@
Hamilton's Principle of Hamiltonian mechanics. Of course, for our %%@
simple systems with non-vanishing Hessian, eq. \ref{nonzerohessian}, %%@
these are equivalent; and so  the discussion applies equally to %%@
Lagrangian and Hamiltonian mechanics. But there will also be a %%@
distinction between the Lagrangian and Hamiltonian approaches which is  %%@
worth noting. 

(2): There are, even within Lagrangian mechanics, several other %%@
variational principles (e.g. principles of least action, least %%@
constraint and least curvature) which I will {\em not} discuss. My %%@
reason for ignoring them is not just lack of space. Also, (i): Broadly %%@
speaking, Hamilton's Principle is more important than them, since in %%@
almost all developments of Lagrangian mechanics it acts as the main %%@
postulate, the other variational principles being deduced from it under %%@
various conditions. (ii): Broadly speaking, the philosophical %%@
discussion in Section \ref{philosophicalpsa} carries over to these %%@
other principles. Or so I contend, without having the space to prove %%@
it!\footnote{I admit that there are many philosophically important %%@
differences between the various principles, including about their modal %%@
involvements. Consider for example the different definitions of %%@
variation used in Hamilton's Principle, Gauss' principle of least %%@
constraint and Euler-Lagrange-Jacobi's principle of least action (cf. %%@
e.g. Lanczos 1986, Chap.s IV.8 and V.4-8). But these differences do not %%@
affect the philosophical  discussion of (Modality;3rd) that follows.}

\subsection{The threat and the answer}\label{philosophicalpsa}
In Section \ref{ssec;threatpsa}, I shall state the threat that a %%@
variational principle like Hamilton's Principle poses; this Section %%@
will be wholly philosophical, involving no technicalities of physics. %%@
Then in Section \ref{ssec;answerpsa}, I shall argue that fortunately, %%@
the threat can be answered: the answer will involve technicalities.

\subsubsection{The threat}\label{ssec;threatpsa}
The threat is that a formulation of an actual law (in this case, the %%@
law of motion of a classical mechanical system)\footnote{Recall from %%@
footnote 9 that my saying `actual' here and elsewhere is not meant to %%@
deny the quantum!} that mentions other possible evolutions of the %%@
system apparently violates the principle  that {\em any} actually true %%@
proposition (not only: any law of nature) should  be made true by %%@
actual facts, i.e. goings-on in the actual world. (So the threat does %%@
not depend on the  evolutions mentioned by the law being contralegal: %%@
what matters is that they are not actual.)

I admit that this principle, often called the {\em truthmaker %%@
principle},  is both vague and disputable. Indeed, this is so, even %%@
apart from disputes about the nature of modality (in particular, the %%@
status of possible worlds). For the terms `true', `proposition' and %%@
`fact' are vague and disputable. In particular, philosophers disagree %%@
about whether ({\em contra} Frege) we need a notion of fact distinct %%@
from (especially: more substantial or ``thicker'' than) the notion of a %%@
true proposition; and even those who accept such facts disagree about %%@
the truthmaker principle thus understood, i.e. about whether every true %%@
proposition is made true by such a fact.

 But I think the principle {\em sounds} right when one first hears it: %%@
(witness the fact that it has been articulated by philosophers working %%@
in different philosophical  traditions---cf. Mulligan et al. (1984)). I %%@
also find that non-philosophers endorse the principle. In particular, %%@
it surely  underlies the point often stressed in physics that a %%@
system's history, for given initial conditions, cannot depend on what %%@
ensemble it is considered to be a member of: (cf. the discussion at the %%@
end of Section \ref{stateworldpsa} of the quantum-classical contrast %%@
concerning whether $S$ is physically real, as shown by its influence on %%@
the system's trajectory). 

So I endorse various philosophers' efforts to articulate a  precise and %%@
true form of the principle; where precision and truth will presumably %%@
require the notion of fact to be not {\em too} ``thick''.  Of course, %%@
controversy continues about how to do this. Here are two examples.  %%@
(1): Assuming that each of a certain collection of propositions $A, B, %%@
\dots$  is made true by such a fact, should we say that the same goes %%@
for their Boolean compounds such as $\neg A, A \wedge B$ and $A \vee %%@
B$; which would amount to admitting negative, conjunctive and %%@
disjunctive facts? (2): Are true propositions  made true not by, or not %%@
just by, a fact---but by an object (i.e. individual, particular) or %%@
objects? Most authors would say that this is at most true of some true %%@
propositions, not all. For there is a mis-match between the Boolean %%@
algebra of propositions, and objects---which do not carry a %%@
corresponding Boolean algebra. Thus suppose we say that $A$ and $B$ are %%@
made true by $a$ and $b$ respectively. If we also believe that any such %%@
objects $a, b$ have a mereological fusion $a + b$, we might say $A %%@
\wedge B$ is made true by $a+b$; but there seems to be no object to %%@
make true a disjunction such as $A \vee \neg B$.

Of course, I do not  have  the space to enter into disputes like those %%@
mentioned: (for recent discussions cf. e.g. Armstrong (1997, Chap. 8), %%@
Mellor (2003)). But fortunately, I do not need to. I can leave the %%@
truthmaker principle vague, mainly because I will need only the fact %%@
that various authors advocate some suitably weak form of it. In fact %%@
Lewis himself is one such author: (so that what follows has a further %%@
{\em ad hominem} interest). The reason I will need only this fact is %%@
that the threat posed to the truthmaker principle  by variational %%@
principles is different from the threats and putative counterexamples %%@
mentioned above; and so far as I know, different from those discussed %%@
in the literature---with one exception.

In fact, the literature discusses two broad kinds of threat:---\\
\indent (A): Threats that, like the problems about Boolean compounds I %%@
mentioned, fall squarely in the  tradition of modern analytic %%@
metaphysics. These threats are often broadly logical, and largely %%@
independent of the subject-matter of the propositions concerned; e.g. %%@
the problem of what, if anything, are the truthmakers of universal %%@
generalizations.\\
\indent (B): Threats based on rejecting the initial idea of a %%@
substantive (``thick'') notion of a truthmaking fact. These may arise %%@
either from holding a general ``minimalism'' about truth, or from %%@
holding that some specific subject-matter, such as ethics or %%@
mathematics, has true propositions without any corresponding ``thick'' %%@
facts. (Of course, a position that went further, and denied that the %%@
subject-matter has true propositions, would be more radical as an %%@
``anti-realism'' or scepticism; but it would  not pose a threat to the %%@
truthmaker principle.)\\
\indent As we shall see, variational principles will differ from both %%@
(A) and (B). And this difference will mean that I can make do without a %%@
precise truthmaker principle---at least in a paper that is a first %%@
foray into analytical mechanics' modal involvements!
  
I said there was one discussion in the literature of a threat to the %%@
truthmaker principle similar to that posed by variational principles. %%@
In fact it is by Lewis himself! He briefly discusses how %%@
counterfactuals, analysed in terms of possible worlds as he proposes, %%@
pose such a threat---to which he then replies. In fact, we will  see in %%@
Section \ref{ssec;answerpsa} that  
a variational principle such as Hamilton's Principle can be regarded as %%@
a long conjunction of Lewisian counterfactuals---and this makes the %%@
threats to the truthmaker principle posed by variational principles and %%@
by counterfactuals closely analogous.

So in the rest of this Section, I propose to report Lewis' discussion %%@
of the threat by counterfactuals, and his reply. But it will help set %%@
the stage for that discussion, to first state two precise forms of the %%@
truthmaker principle, as formulated by Lewis. The first illustrates how %%@
the principle can be formulated so weakly as not be threatened (by %%@
variational principles, counterfactuals or indeed any proposition). The %%@
second formulation is stronger, and {\em is} threatened by variational %%@
principles and counterfactuals: (it will also clarify how the threat %%@
posed by these is different from those in the %%@
literature).\footnote{Both formulations also illustrate how in some of %%@
his work, Lewis  incorporates current positions and insights into his %%@
own philosophical  system---and in doing so, often makes them more %%@
vivid and precise. Indeed, this ability to  incorporate  ideas that are %%@
``in the air'' into his system, and to make them vivid and precise, is %%@
one of his great strengths as a philosopher. His writings provide many %%@
examples: e.g. his treatment of indexicality as attitudes {\em de se} %%@
(1979), and his treatment of chance (1980, 1994).} 

(1): {\em Truth supervenes on being}: Bigelow (1988, 132-3, 158-9) %%@
makes the idea of truthmakers more precise along the following lines: %%@
that how things are determines which propositions are true---which he %%@
expresses with the slogan `truth supervenes on being'. Lewis %%@
incorporates this in his framework of possible worlds, in such a way %%@
that it becomes {\em a priori}. Accordingly, Lewisian counterfactuals %%@
and variational principles---as well as the other cases considered in %%@
the literature---pose no threat to it. (As I said, some forms of the %%@
truthmaker principle are weak!) Thus Lewis  uses his ideas (1988) %%@
that:\\
\indent (a): a proposition is a set of worlds, viz. the worlds where %%@
the proposition  is true;\\
\indent (b): a subject-matter is a partition of the set of all worlds, %%@
with any two worlds in a cell of the partition matching as regards the %%@
subject-matter; and\\
\indent (c): a proposition is wholly about a subject-matter if it (i.e. %%@
its set of worlds) is a union of the cells of the subject-matter's %%@
partition.\\
\indent Lewis then construes Bigelow's `being' as the largest %%@
subject-matter, i.e. the maximal partition. So truth's supervening on %%@
being becomes an {\em a priori} truth. Every proposition is a union of %%@
the cells of the maximal partition; and which of those cells  contains %%@
the actual world trivially determines which propositions are actually %%@
true (1992, Section 6; 1994, Section 1; 2003, sections %%@
1-2).\footnote{Cf. Lewis (2003) for discussion of further aspects. In %%@
particular: (a) this notion of aboutness does not suit necessary or %%@
impossible propositions---it is intensional but not hyperintensional; %%@
(b) cells of the maximal partition might not be singleton sets of %%@
worlds, since there might be indiscernible worlds and we might ban %%@
non-qualitative propositions.}
 
\indent (2): {\em Counterparts as truthmakers}: Lewis has recently %%@
proposed that some propositions have truthmakers that are objects---in %%@
his jargon: individuals; (2003, Section 3 et seq., overcoming previous %%@
scepticism in e.g. his (1992, Section 5)). More precisely, he defines a %%@
possible individual $a$ to be a truthmaker for a proposition $A$ iff %%@
every world where $a$ exists is a world where $A$ is %%@
true.\footnote{Others give verbally the same definition, though in %%@
their own metaphysical frameworks: e.g. Armstrong (1997, p.115), %%@
Bigelow (1988, p.122,126).} Here `every world where $a$ exists' must be %%@
understood, in the light of Lewis' denial of strict identity across %%@
possible worlds, in terms of his counterpart theory. Lewis goes on to %%@
show that counterpart theory yields a truthmaker in his precise sense %%@
for many propositions, in particular for predications. Besides, a %%@
postscript (co-authored with G. Rosen) argues that the proposal can be %%@
extended to other propositions, even negative existentials like `There %%@
are no unicorns'.  (Lewis also compares his proposal with proposals for %%@
facts as truthmakers made by Armstrong and Mellor.) Again, I cannot go %%@
into details. For this paper's purposes, it suffices to note the %%@
contrast with (1). That is to say: for objects as truthmakers, the %%@
threat that concerns us arises again: variational principles and %%@
Lewisian counterfactuals, with their transworld comparisons, apparently %%@
do not have this sort of truthmaker.

These items (1) and (2) set the stage for Lewis' discussion of how %%@
counterfactuals threaten the idea of truthmakers. For that discussion %%@
falls between (1) and (2), in the sense that there {\em is} a threat %%@
(like (2) and unlike (1)), but one that (he maintains) can be answered %%@
(unlike (2)). I turn to reporting that discussion. 

Lewis of course  recognizes that his proposed truth-conditions for %%@
counterfactuals in terms of similarity between possible worlds threaten %%@
the the idea of truthmakers; (although his discussion does not use the %%@
term `truthmaker', the connection will be clear). After all, Lewis %%@
proposes for an actually true counterfactual, truth-conditions in terms %%@
of other worlds! Thus recall that, roughly speaking, $A \rightarrow C$ %%@
is actually true if some $(A \& C)$-world is closer (i.e. more similar) %%@
to the actual world than any $(A \& \neg C)$-world is. So he writes:
\begin{quote}
Here is our world, which has a certain qualitative character. (In as %%@
broad a sense of `qualitative' as may be required---include irreducible %%@
causal relations, laws, chances, and whatnot if you believe in %%@
them.)\footnote{[By JNB]: Thus this threat is independent of Lewis' %%@
neo-Humean analyses of causation, law and chance; as also of his more %%@
speculative additional doctrine, Humean supervenience.} There are all %%@
the various $A$-worlds, with their various characters. Some of them are %%@
closer to our world than others. If some $(A \& C)$-world is closer to %%@
our world than any $(A \& \neg C)$-world is, that's what  makes the %%@
counterfactual true at our world. Now ... it's the character of our %%@
world that makes some $A$-worlds be closer to it than others. So, after %%@
all, it's the character of our world that makes the counterfactual %%@
true---in which case why bring the other worlds into the story at all? 

To which I reply that it is indeed the character of our world that %%@
makes the counterfactual  true. But it is only by bringing the other %%@
worlds into the story that we can say in any concise way what character %%@
it takes to make what counterfactuals true. The other worlds provide a %%@
frame of reference whereby we can characterize our world. By placing %%@
our world within this frame, we can say just as much about its %%@
character as is relevant to the truth of a counterfactual (1986, p. %%@
22).
\end{quote}

This passage makes two main claims, one in each paragraph:\\
\indent (Actual): although Lewis' truth-conditions mention other %%@
worlds, it is the character of the actual world that makes the %%@
counterfactual actually true;\\
\indent (Concise): mentioning other worlds is the only concise way to %%@
state what in the actual world's character is relevant to the %%@
counterfactual's truth.\\
 Of these two claims, (Actual) is the more important for us---it %%@
summarizes both the threat to truthmakers and Lewis' reply. But I shall %%@
also briefly discuss (Concise).  

We can better understand (Actual) by recalling Lewis' (1986, p. 62) %%@
distinction between (a) relations that supervene on the intrinsic %%@
properties of their {\em relata}, which Lewis calls `internal', and (b) %%@
relations that do not thus supervene, which I will call `external'. (I %%@
will not need Lewis' doctrines about which properties are intrinsic, %%@
and can make do with some intuitive if disputable examples of intrinsic %%@
properties. Nor will I need Lewis' allowance that a relation might %%@
supervene on the composite of the {\em relata} taken together: his main %%@
example of this category being spatiotemporal relations.) Thus %%@
relations of similarity or difference in intrinsic respects are %%@
internal; so that if an object's mass is an intrinsic property of it, %%@
the relation `is more massive than' is internal. An example of an %%@
external relation would be `has the same owner as': $a$ and $a'$ could %%@
match in all their intrinsic properties and yet a person might own $a$ %%@
and some other object $b$, but not $a'$; so that `has the same owner %%@
as' holds of $\langle a, b \rangle$ but not $\langle a', b \rangle$.

Lewis applies this distinction not just to relations between objects in %%@
a single world, but to objects in different worlds. Thus a sentence %%@
such as `He is slimmer than he would have been without the diet' %%@
reports an internal relation between objects in different worlds (a man %%@
and one of his counterparts). A sentence reporting a transworld %%@
external relation seems harder to construct; I suppose because our %%@
thought and language has little use for them. But Lewis' own %%@
counterpart theory  gives examples. For counterparthood, though it %%@
sometimes emphasises intrinsic similarity, often emphasises extrinsic %%@
similarity, especially as regards the object's origins (Lewis 1986, %%@
p.88). Thus two objects $a$ and $a'$ (in the same world, or in two %%@
different worlds) might be duplicates, while only $a$ is a counterpart %%@
of some object $b$ in another world---say an actual object %%@
$b$.\footnote{Here is an example with $a, b$ both actual, and indeed %%@
identical: `an atom-for-atom replica of Humphrey (as he actually was %%@
at, say, noon 4 July 1968), who had been born of different parents than %%@
the actual Humphrey (in say Latvia, never setting foot in the USA %%@
etc.), would not have been [folk-language, according to Lewis, for: %%@
would not have been a counterpart of] Humphrey'. Here, $a = b =$ the %%@
actual Humphrey, and $a' =$ the replica.  Another example, with $a$ and %%@
$b$ in different worlds: `Each of two people might be atom-for-atom %%@
replicas of Humphrey as he actually was at noon, 4 July 1968; but only %%@
the person whose origin matched (at least: sufficiently closely) that %%@
of the actual Humphrey, would be Humphrey'. Here, $a, a'$ are the %%@
replicas, $b$ is the actual Humphrey.}

Furthermore, Lewis also takes worlds to be objects (in short: the %%@
mereological fusion of their parts) and so allows them as {\em relata}; %%@
and therefore applies this distinction to relations between worlds. And %%@
he says explicitly (1986, p. 62,177) that since the relation of %%@
closeness between possible worlds used in his analysis of %%@
counterfactuals is a relation of similarity, it is internal. Hence his %%@
claim in (Actual) that the truth-values of counterfactuals are %%@
determined by the character of our world. For the character of our %%@
world  determines which worlds are similar to it. (Though it  is a %%@
vague and controversial matter which respects of similarity are %%@
relevant to the truth-conditions of counterfactuals  (Lewis 1979a), any %%@
resolution of those issues will render the overall similarity relation %%@
internal.)

The connection of Lewis' (Actual) with the idea of truthmakers is %%@
clear. Though `truthmaker' is a philosophical term-of-art awaiting %%@
strict definition, the way that Lewis' truth-conditions mention other %%@
worlds makes one think that---whether one takes truthmakers to be facts %%@
or objects---a counterfactual has truthmakers ``scattered across the %%@
worlds'': apparently violating the principle that actual truths have %%@
actual truthmakers. To which threat, Lewis replies: `No worries: {\em %%@
which} facts, objects etc. in other worlds get mentioned in the %%@
truth-conditions is wholly determined by the character of the actual %%@
world---and that is sufficient for satisfying the idea that actual %%@
truths have actual truthmakers.' And Lewis might well go on: `If you %%@
want, you can call the facts, objects etc. in the other worlds that get %%@
mentioned in the truth-conditions  `truth-makers'. But the point %%@
remains that  their being scattered across the worlds is innocuous. The %%@
fact that the character of the actual world determines them (and %%@
thereby the truth-value of the counterfactual) is sufficient to satisfy %%@
the spirit, if not the letter, of the truthmaker principle that `actual %%@
truths have actual truthmakers'.' 

So much by way or explicating Lewis' claim (Actual), i.e. his reply to %%@
the threat posed by counterfactuals. I think that within the framework %%@
of Lewis' metaphysics, it faces no objections. But of course, my main %%@
purpose is not to report or defend Lewis. Rather, the point of my %%@
endorsement of (Actual) is that, as we shall see in Section %%@
\ref{ssec;answerpsa}, variational principles can be similarly %%@
reconciled with the spirit, if not the letter, of the principle  %%@
`actual truths have actual truthmakers'---just because we can read such %%@
principles as long conjunctions of Lewisian counterfactuals.

I turn to briefly discuss Lewis' claim (Concise): that mentioning other %%@
worlds is the only concise way to state what in the actual world's %%@
character is relevant to a counterfactual's truth. I would have liked %%@
Lewis to say more about this, especially in view of (a) the importance %%@
in his philosophical system of the threatening counterfactuals, and (b) %%@
the importance in his late work of the threatened idea of truthmakers. %%@
One naturally asks: {\em why} is mention of other worlds the only %%@
concise way to describe the relevant part of the the actual world's %%@
character? But so far as I know, this passage is all Lewis says on the %%@
topic.\footnote{Here is an analogy that I  use in explaining Lewis' %%@
reply. To describe Buenos Aires concisely to a friend who is unfamiliar %%@
with it, you might forego listing its intrinsic properties, and instead %%@
use a comparison with something familiar to your friend; thinking of %%@
the harbour and summer in January, you might say, for example, `It's %%@
like a Spanish-speaking Sydney'. I confess I believed Lewis invented %%@
this (Australophile!) analogy; but I cannot find it in his published %%@
work---maybe he just told me it. Alan Hajek tells me that he, Hajek, %%@
invented this same analogy for the same purpose, except that Hajek's %%@
example was that Perth is the Australian city most similar to San %%@
Diego. Hajek also reports that Lewis endorsed the analogy; (so maybe %%@
Lewis got the analogy from Hajek, and passed it to me). But more %%@
important than who invented it: according to Hajek, Lewis insisted that %%@
it was a fact {\em about San Diego} that the Australian city most %%@
similar to it is Perth---cf. Lewis' claim (Actual) above.}

In any case, we will see in Sections \ref{ssec;answerpsa} and %%@
\ref{cvrevisitedpsa} a contrast between Lewis' philosophical system and %%@
our concern: mechanics and the calculus of variations. In our simpler %%@
and more technical framework, one {\em can} say more about why %%@
mentioning possible histories is useful. Indeed, one can also say more %%@
about the analogue of Lewis' claim (Actual): about the character of the %%@
actual world (i.e. the this-worldly truthmakers) that makes a %%@
variational  principle true.
  
Finally, before turning to this---i.e. answering the threat that %%@
variational principles pose to the idea of truthmakers---I should set %%@
aside two ways in which a philosopher might dismiss this threat, so %%@
that there would be no ``case to answer''.

 (1): The first dismissal echoes (B) above: i.e. the rejection of the %%@
idea of substantive, truthmaking, facts, either generally or for a %%@
particular subject matter. Thus a philosopher might think variational %%@
principles are not ``in the market'' for truthmakers, for a variety of %%@
reasons: ranging from\\
\indent (i): ``minimalism'' about truth (either generally or for %%@
variational principles); through\\
\indent (ii): some kind of instrumentalism (i.e. denial that %%@
variational principles are  true, and even that they purport to be %%@
true, yet acceptance of their usefulness); to\\
\indent (iii) some kind of eliminativist ``anti-realism'' (i.e. denial %%@
of usefulness as well as truth).

Needless to say, I will not try to reply at length to all these %%@
positions! Suffice it to make one reply to each of (i)-(iii).\\
\indent (i'): I am unconvinced of minimalism in general, and see no %%@
special motivation for holding it  for variational principles.\\
\indent (ii'): I have two replies to instrumentalism: the first general %%@
and firm, the second special and yielding. (a): First, I see no %%@
motivation for  instrumentalism about all variational principles, %%@
except as an instance of a general instrumentalism about all %%@
theoretical claims: a general instrumentalism which I have no truck %%@
with, and which is anyway nowadays unpopular, displaced in large part %%@
by van Fraassen's constructive empiricism.\footnote{But there is a rich %%@
subject here. St\"{o}ltzner (2003) is a fascinating study of the %%@
logical empiricists' treatment---and mistreatment!---of variational %%@
principles in mechanics.} (b): However, I will concede at the end of  %%@
Section \ref{ssec;answerpsa} that instrumentalism about variational %%@
principles {\em is} tenable for non-simple mechanical systems.\\
\indent (iii'): I wholly reject the idea that variational principles %%@
are not useful: I shall develop this theme in Section %%@
\ref{cvrevisitedpsa}.   

(2): Finally, a philosopher might say that the variations mentioned in %%@
variational principles have nothing to do with possibilities of the %%@
sort discussed in the literature about truthmakers (or in modal %%@
metaphysics generally). Again, I have no truck with this. As I said in %%@
Section \ref{Threepsa}, mechanics is up to its ears in modality, of %%@
{\em some} kind or kinds. And no sign is ever given that modal %%@
locutions like `could', with which the notions and mathematics of %%@
virtual displacements, variations etc. are introduced, are to be %%@
understood differently from elsewhere in science or everyday life. So %%@
why should they be?

 \subsubsection{The answer}\label{ssec;answerpsa}
To answer the threat, I shall adapt a two-stage strategy which is %%@
straightforward, and traditional in philosophy. According to this %%@
strategy, when one is confronted with apparently problematic entities, %%@
one has to consider two tasks. The first takes one of two forms; but it %%@
is compulsory, in the sense that one must undertake either the first %%@
form or the second. On the other hand, one faces the second task only %%@
if one undertakes the second form of the first task; and even then, the %%@
second task is  optional, not compulsory---though  succeeding in it %%@
would be a significant merit of one's philosophical  position.

Thus the first task is as follows. One must either show that the %%@
problem is an illusion---the entities are not really problematic, after %%@
all: they can be vindicated. I shall call this task (Vindicate). Or, %%@
accepting that the entities really are problematic, one must show how %%@
to do without them: one must eliminate them. I call this (Eliminate). %%@
If one undertakes (Eliminate), one should, if  possible, undertake the %%@
second task: to show how it is useful or convenient to speak as if the %%@
entities exist (so as to explain, perhaps even justify, ``how the folk %%@
speak''). I shall call this task (Useful).

For variational principles in mechanics, the entities at issue---the %%@
possible histories of the system---are not themselves problematic; (I %%@
of course set aside the debate about the nature of possibilities, i.e. %%@
Lewis' modal realism vs. various ersatzisms and fictionalisms). Rather, %%@
what seems problematic is the role these entities are assigned: viz. %%@
being, when taken together with the actual history, truthmakers of %%@
actual truths (indeed the actual laws).  But clearly, we can adapt the %%@
two-stage strategy to apparently problematic roles rather than %%@
entities. Thus we envisage arguing that:\\
\indent (Vindicate): This role of possible (indeed, contralegal) %%@
histories can be vindicated---it is not problematic, after all; or %%@
instead that\\
\indent (Eliminate): This role of possible histories can be %%@
eliminated---the laws can be formulated without invoking it; in which %%@
case we should also try to argue that\\
\indent (Useful): The variational formulation of the laws is %%@
nevertheless useful, or even advantageous compared with  formulations %%@
that do not mention possible histories. (Cf. (iii') at the end of %%@
Section \ref{ssec;threatpsa}.)\\
\indent In this Section, I will discuss (Vindicate) and (Eliminate), in %%@
order. But I postpone (Useful) to Section 5.2.

\paragraph{5.1.2.A Vindicating possible histories}
One can argue for (Vindicate) on strict analogy with Lewis' own answer %%@
to the threat that counterfactuals apparently pose to the truthmaker %%@
principle: i.e. Lewis' claim (Actual), reported in Section %%@
\ref{ssec;threatpsa}. For there is a striking analogy between what a %%@
variational principle says and Lewis' proposed truth-conditions for %%@
counterfactuals $A \rightarrow C$. Roughly speaking, a history of the %%@
mechanical system corresponds to a Lewisian possible world (cf. Section %%@
\ref{stateworldpsa}'s suggestion); and similarity of histories is a %%@
matter of first fixing the configurations at the end-points, and %%@
secondly closeness of the values of $q$ and ${\dot q}$. Given this %%@
correspondence, a variational principle turns out to be an infinite %%@
conjunction of Lewisian counterfactuals. 

To spell this out, let us first recall Lewis' proposal (cf. Section %%@
\ref{configworldpsa}). To avoid the Limit Assumption (that there is a %%@
set of $A$-worlds all tied for first equal as regards similarity to the %%@
actual world @), and setting aside the case of vacuous truth, Lewis %%@
proposes that a counterfactual $A \rightarrow C$ is true at @ iff: some %%@
sphere $S$ in the system $\$_{@}$ contains at least one $A$-world, and %%@
$A \supset C$ is true at every world in $S$; (i.e. $C$ is true at every %%@
$A$-world in $S$).

Turning to variational principles, I shall take Hamilton's Principle; %%@
though essentially the same analogy could be drawn with any number of %%@
principles. The principle says that the actual history is a stationary %%@
point of the action. Here, `stationary point' rather than `minimum' %%@
allows that:\\
\indent\indent (i): the actual history could be a maximum of the action %%@
$ \int L \; dt$, not a minimum;\\
\indent\indent (ii): the minimum or maximum need only be local;\\
\indent\indent (iii): the actual history could be a point of inflexion %%@
(associated with the vanishing of second derivative of the action, not %%@
just the first).

But I now need to spell this out in more detail than I did in Section %%@
\ref{ssec;cvreview} (paragraph 3). Roughly speaking, Hamilton's %%@
Principle says that:
\begin{quote}
For any one-parameter family, parametrized by $\al$ say, of %%@
kinematically possible histories of the mechanical  system, that may %%@
deviate from the actual history between $t_0$ and $t_1$, but must match %%@
the actual history as regards the configurations $q_0, q_1$ at times %%@
$t_0,t_1$:\\
the action as a function of $\al$, $I(\al) = \int^{t_1}_{t_0} L \; dt$ %%@
with the integral  taken along the history labelled by parameter-value %%@
$\al$, has zero gradient at the value of $\al$ corresponding to the %%@
actual history.
\end{quote}
To be precise, we consider any one-parameter family, parametrized by %%@
$\al$, of curves from $q_0,t_0$ to $q_1,t_1$; so we write $q_i = %%@
q_i(t,\al)$. We also suppose that the curve (let us call it @!) that %%@
makes stationary the integral  $I(\al) = \int^{t_1}_{t_0} L \; dt$ %%@
(taken along the curve labelled by $\al$) has parameter-value $\al = %%@
0$; which we write as $q_i(t) := q_i(t,0)$. So the family of curves is %%@
given by $q_i(t,\al) = q_i(t) + \al \eta_i(t)$. Then for the action %%@
integral to be stationary at $\al = 0$ means that in the Taylor %%@
expansion about $\al = 0$, i.e.
\be
I(\al) \equiv \int^{t_1}_{t_0} L(q_i(t) + \al \eta_i(t), {\dot q}_i(t) %%@
+ \al {\dot \eta}_i(t), t) \; dt = I(0) + \al\left(\frac{\pl I}{\pl %%@
\al}\right)_{\al = 0} + {\rm O}(\al^2) \; ,
\label{definestationarity}
\ee
we have: 
\be
\left(\frac{\pl I}{\pl \al}\right)_{\al = 0} = \; 0 \; .
\label{definestatyy2}
\ee
That is, by the elementary definition of the derivative as a limit of a %%@
quotient:
\be
\forall \varepsilon > 0 \;\; \exists \dd > 0 \;\; \forall \; 0 < \al %%@
\leq \dd \; , \;\;\; \frac{I(\al) - I(0)}{\al} < \varepsilon \;\; .
\label{elyderivativezero}
\ee

Now I can state the analogy between Lewis' truth-condition and %%@
Hamilton's Principle (for our given one-parameter family of curves). We %%@
take the parameter $\al$ to define a system of nested spheres (sets of %%@
curves) $\$_{@}$, centred on the curve @ which itself has parameter %%@
$\al = 0$: the spheres are defined by inequalities $\al \leq r \in %%@
\mathR$, so that a smaller value of $\al$ represents greater similarity %%@
to @. 

With this understanding, we can read Hamilton's Principle, in the form  %%@
eq. \ref{elyderivativezero}, as a battery of Lewisian counterfactuals; %%@
indeed as a infinitely long conjunction of them (there are at least %%@
continuously many conjuncts).  This plethora of counterfactuals arises %%@
from two sources:\\
\indent (i): the continuously large range of values of $\varepsilon$; %%@
and\\
\indent (ii): for given $\varepsilon$, the at-least-continuously large %%@
range of antecedents $A$ that are false at the actual curve @ but true %%@
at some curve in the family with a parameter-value $0 < \al \leq \dd$. %%@
(Here we think of $\dd$ as determined using eq. \ref{elyderivativezero} %%@
from the given $\varepsilon$.)\\
But on the other hand, the counterfactuals are similar as regards their %%@
consequents $C$: they are all given by the inequality in eq. %%@
\ref{elyderivativezero}.\footnote{Besides, there is another source of %%@
yet more counterfactuals, viz. the various choices for our %%@
one-parameter family of curves: which I have for the sake of clarity %%@
suppressed, by fixing on a single family.}

That is to say: we can read eq. \ref{elyderivativezero} as saying that %%@
for each value of $\varepsilon$, and any $A$ that is false at the %%@
actual curve @ but true at some curve in the family with a %%@
parameter-value $0 < \al \leq \dd \equiv \dd(\varepsilon)$: that curve %%@
labelled by $\al $ makes the quotient $\frac{I(\al) - I(0)}{\al}$ less %%@
than $\varepsilon$, as do all curves with an $\al$ in the same range.

 In other words, now using the Lewisian spheres defined by inequalities %%@
$\al \leq r \in \mathR$, so as to talk about counterfactuals:---
\begin{quote} 
for all $\varepsilon$, and all such $A$ (so that $\varepsilon$ fixes a %%@
range of $\al$, viz. $0 < \al \leq \dd \equiv \dd(\varepsilon)$, and  %%@
$A$ is true at such an $\al$, but false at $\al =0$): the Lewisian %%@
counterfactual
\begin{quote}
$A \rightarrow$ [the quotient $\frac{I(\al) - I(0)}{\al}$ is less than %%@
$\varepsilon$]
\end{quote} 
is true at @.
\end{quote}
(Incidentally: we need $A$ to be false at @, i.e. the counterfactual $A %%@
\rightarrow C$ to be contrary to fact, so as to force $0 < \al$, so %%@
that the quotient is well-defined.) To sum up: Hamilton's Principle is %%@
an infinitely long conjunction of Lewisian counterfactuals.

Returning (at last!) to philosophy: this discussion makes clear the  %%@
analogy with Lewis' claim (Actual), reported in Section %%@
\ref{ssec;threatpsa}, and thereby our argument for (Vindicate). For all %%@
the ingredients in the above transworld comparisons involve only %%@
internal relations, either between objects in possible worlds (i.e. %%@
components of the system in possible histories, possible curves) or %%@
between worlds (histories, curves) themselves. Thus similarity between %%@
worlds is a matter of: first, shared initial and final configurations; %%@
and second, closeness of values of $q$ and so $\dot q$, where closeness %%@
is encoded by the parameter $\al$. So the character of the actual world %%@
@ (i.e. the course of values of the functions $q, {\dot q}$ along the %%@
actual history) determines, for any antecedent proposition $A$, at %%@
which if any of the histories in the system of nested spheres (sets of %%@
histories) $\$_{@}$, $A$ is true. The case is even clearer as regards %%@
the consequent $C$, which as noted is similar for all the %%@
counterfactuals involved. The inequality $\frac{I(\al) - I(0)}{\al} < %%@
\varepsilon$ compares the values of $I$ on different histories. This is %%@
an internal relation between worlds, i.e. histories, since whether or %%@
not this relation holds is determined by the values $I(\al)$ and %%@
$I(0)$: which are part of the intrinsic natures of the worlds.  To sum %%@
up, by echoing Lewis' claim (Actual): you can say if you like that the %%@
truthmakers of Hamilton's Principle are ``scattered across the %%@
worlds''; but the spirit, if not the letter, of the truthmaker %%@
principle is satisfied.\footnote{To make the argument clearer, I have %%@
suppressed a wrinkle about spatiotemporal relations. According to %%@
Lewis, these relations are not internal: they supervene on the %%@
intrinsic properties of the composite of the relata, not on the %%@
properties of the relata individually. This might seem an obstacle to %%@
my argument, since   spatiotemporal relations are surely involved in %%@
assessing similarity of worlds in our sense, viz. courses of values of %%@
$q$ and $\dot q$. But no worries. It is as relations between objects in %%@
a single world that spatiotemporal relations are not internal; but for %%@
variational principles, the assessment of similarity makes a comparison %%@
of the spatiotemporal structures of entire worlds---and the ensuing %%@
similarity is obviously an internal relation between the worlds.}

So much for arguing for (Vindicate). But notwithstanding this success, %%@
some  will still worry! Some philosophers are very wary about modality. %%@
And even if one relishes modality, it may seem risky to rely on %%@
satisfying just the spirit, and not the letter, of the truthmaker %%@
principle: especially while a precise and correct  formulation of the %%@
principle remains controversial---for the formulation eventually agreed %%@
on might have a more demanding spirit than one now guesses! So I ought %%@
also to consider how one might argue for the other %%@
option---(Eliminate). 

\paragraph{5.1.2.B Eliminating possible histories}
Focussing again on Hamilton's Principle, I shall first consider whether %%@
there is a statement (or statements) equivalent to Hamilton's %%@
Principle, that does not mention possible histories of the system. Here %%@
`equivalence' means logical equivalence; or perhaps mathematical %%@
equivalence, understood in the usual way as logical equivalence, once %%@
given the assumption of appropriate pure mathematical propositions. %%@
However, the idea of (Eliminate) does not require that there be such an %%@
equivalence: it would surely be enough that there are alternatives to %%@
Hamilton's Principle---i.e. statement(s) that  do not mention possible %%@
histories, and yet function as laws of motion. So after considering %%@
equivalence, I shall discuss this alternative.

 For the simple mechanical systems we have focussed on since Section %%@
\ref{ssec;cvreview}, there {\em are} equivalent statements. For as %%@
noted in Section \ref{ssec;cvreview}, the Lagrange equations eq. %%@
\ref{elpara=t} are, for simple systems, not only necessary but also %%@
{\em sufficient} for Hamilton's Principle eq. \ref{hpsimple}. And these %%@
equations do not mention possible histories. Agreed, they are modally %%@
involved; at least if we take them as putative laws of analytical %%@
mechanics---as we no doubt should! For then we will take them as %%@
applying to possible as well as actual initial conditions (given by the %%@
$q$s and $\dot q$s), and to possible as well as actual problems. That %%@
is, they will illustrate (Modality;1st) and (Modality;2nd), in Section %%@
\ref{Threepsa}'s classification; but not (Modality;3rd). In short: %%@
Hamilton's Principle can be regarded for simple systems as a corollary %%@
of the ``kosher'' this-worldly laws, Lagrange's equations eq. %%@
\ref{elpara=t}. 

The same point applies to Hamiltonian mechanics for simple systems with %%@
non-zero Hessian, eq. \ref{nonzerohessian}. In this context, Hamilton's %%@
equations eq. \ref{canleqnswithinCV2nd} are equivalent, by the Legendre %%@
transformation eq. \ref{legtrfmnsymmicform2nd}, to Lagrange's equations %%@
eq. \ref{elpara=t}. So again, taking Hamilton's equations as laws of %%@
analytical mechanics---as we no doubt should---they illustrate %%@
(Modality;1st) and (Modality;2nd) but not (Modality;3rd).

Incidentally, Hamiltonian mechanics raises another point, concerning %%@
the free variation of the $p$s in the modified Hamilton's Principle eq. %%@
\ref{canlvarnprobwithinCV2nd}. This gives another illustration of %%@
(Modality;3rd). But it is a more ``extreme'' illustration than that %%@
given by the original Hamilton's Principle eq. \ref{hpsimple}. For the %%@
latter, we contralegally vary $q$ and so $\dot q$. But for the modified %%@
Hamilton's Principle, once we are given such a variation of $q$ (and so %%@
$\dot q$), we independently vary the $p$s (violating $p = {\pl L}/{\pl %%@
{\dot q}}$). So our variations are ``doubly contralegal''. 

But what about arguing for (Eliminate) for mechanical  systems that are %%@
{\em not} simple: where the above equivalence breaks down? That is (cf. %%@
discussion after eq. \ref{hpsimple}): for systems for which the %%@
this-worldly Lagrange equations are only necessary but not sufficient %%@
for Hamilton's Principle? 

In fact, I believe (Eliminate) can be defended for such systems; (and %%@
more generally, for systems for which the this-worldly dynamical %%@
equations are only necessary but not sufficient for a variational %%@
principle). I must postpone this topic to another occasion; not least %%@
because it has technical aspects (cf. Papastavridis 2002, pp. 960-973), %%@
as well as philosophical ones. But I should  admit here that this %%@
defence opens the door to instrumentalism about variational principles. 
Thus suppose one says that the laws of motion are  given by the (true %%@
and this-worldly!) Lagrange equations, not by Hamilton's principle. %%@
Then it seems one can  turn instrumentalist about the latter: since %%@
these principles  are sufficient but not necessary  for the laws, one %%@
need not accept them as true, in order to agree that they have various %%@
uses. And they certainly {\em do} have uses numerous and important %%@
enough to earn them their central place in expositions of mechanics, %%@
even if they are ``merely'' instruments. I will return to this in %%@
Section \ref{cvrevisitedpsa}'s defence of (Useful). For the moment, I %%@
just mention one main use: a variational principle is often used as a %%@
way of guessing or deriving the laws of motion, since it is often %%@
easier to guess a Lagrangian that obeys a required symmetry than a set %%@
of laws of motion that obeys it.

Finally, a philosophical point that bring us back to Lewis. There is a %%@
temptation to say that a mystery remains, even {\em after} the argument %%@
for (Eliminate) for simple systems. It is tempting to ask:
how can one of two {\em equivalent} formulations of a law (or %%@
theory)---Hamilton's Principle, on the one hand, and Lagrange's %%@
equations or Hamilton's equations, on the other---have a modal %%@
involvement that the other lacks? Indeed, 
more generally: How can there be a logical equivalence between a %%@
proposition with ``this-worldly truth-conditions'' and one making %%@
``transworld comparisons''?\footnote{Most philosophers agree that there %%@
may well be a notion of theoretical, not merely empirical, equivalence %%@
such that laws or even whole theories  that are theoretically %%@
equivalent  could yet have heuristic, and even ontological, %%@
differences. Still, there can seem to be a mystery about our argument %%@
for (Eliminate); since both the equivalence of the formulations, and %%@
their having different modal involvements, seem matters of logic.}

I think Section \ref{ssec;threatpsa}'s discussion, especially Lewis' %%@
claim (Actual), gives most of the reply to this question; and I will %%@
not rehearse it again. But another point, {\em independent} of (Actual) %%@
and indeed of the general idea of truthmakers,  is worth stressing.
Namely, there is no logical or semantical problem about evaluating as %%@
true at a single world a proposition making a transworld comparison. %%@
After all, this is exactly what is proposed by analyses of %%@
counterfactuals like those of Lewis and Stalnaker; and  proposed by %%@
these analyses as  semantical doctrines, independently of Lewis' %%@
metaphysical thesis (Actual). So suppose someone thought some %%@
propositions make transworld comparisons, in the strong sense that %%@
their truth-conditions (or if you like: truthmakers) are scattered %%@
across the worlds in ways {\em not} determined by internal relations of %%@
those worlds to the actual world. Such a person could nevertheless %%@
accept, as a matter of logic or semantics, that such a proposition be %%@
assigned a truth-value relative to the actual %%@
world.\footnote{Incidentally, the use of truth-assignments relative to %%@
two or more worlds in many-dimensional modal logic (and similarly: %%@
relative to two or more times in temporal logic) is, so far as I can %%@
tell, no evidence against the truthmaker principle. For these logics %%@
invoke multiple worlds or times to keep track of rigidified uses of %%@
`actually' or temporal indexicals: not to keep track of truthmakers %%@
scattered among the worlds---indeed, not even `scattered among the %%@
worlds' in the innocuous sense allowed by Lewis' (Actual), i.e. in the %%@
innocuous sense of determined by internal relations among the worlds.}  

To sum up this Section:--- I first argued for (Vindicate). Variational %%@
principles' mention of possible histories can be vindicated by an %%@
argument parallel to Lewis' argument that counterfactuals are made true %%@
by the character of the actual world---since their mention of other %%@
worlds reflects only internal relations between worlds. This parallel %%@
was based on showing that a variational  principle is itself an %%@
infinite conjunction of Lewisian counterfactuals. Then I argued for %%@
(Eliminate), at least for simple systems. That is: we can identify the %%@
this-worldly truthmakers of Hamilton's Principle, namely {\em via} %%@
Lagrange's (or equivalently: Hamilton's) equations. 

\subsection{The uses of variational principles}\label{cvrevisitedpsa}
I turn to the claim that at the start of Section \ref{ssec;answerpsa}, %%@
I dubbed (Useful): that formulating classical dynamical laws as %%@
variational principles is useful, or even advantageous compared with %%@
other formulations.

I admit that I shall duck out of giving a general argument for %%@
(Useful). Rather like Lewis with his claim (Concise) about %%@
counterfactuals (cf. Section \ref{ssec;threatpsa}), I offer no single %%@
general advantage of variational formulations. My reason is that the %%@
advantages are many, diverse and sometimes very technical. The calculus %%@
of variations remains an active research area,  with deep connections %%@
to various branches of mathematics in addition to mechanics. (For a %%@
taster, cf. e.g. Courant and Hilbert (1953, Chap IV); for a banquet, %%@
cf. Giaquinta and Hildebrandt (1996).) So it would be well-nigh %%@
impossible even to list, let alone discuss, the advantages gained by %%@
adopting the notions, and general perspective, of the calculus of %%@
variations; not just for mechanics, but for any field that uses %%@
variational principles. So I shall just mention two examples of %%@
advantages of variational principles in analytical mechanics that would %%@
appear on any such list.\\
\indent (i): The role of variational principles in understanding %%@
symmetry; especially the way that symmetries of the Lagrangian give %%@
Noether's theorems.\\
\indent (ii): I choose my second example to illustrate how a piece of %%@
formalism within a theory can be advantageous not only as regards that  %%@
theory, but also in illuminating another theory; (and maybe even %%@
heuristically useful in constructing that other theory). I have in mind %%@
how Hamilton's Principle illuminates the path integral formulation of %%@
quantum theory; both by providing a classical limit of it, and by %%@
heuristically suggesting it.

Finally, I should note an important topic related to (Useful): the %%@
topic, not of the {\em advantages} of a variational formulation of %%@
laws, but of the {\em conditions} under which such a formulation can be %%@
given. This is a large topic, which has been investigated since the %%@
nineteeth century, mostly in the more precise form: what are the %%@
necessary and sufficient conditions for a set of dynamical %%@
(differential) equations governing variables $q_i$ to be the %%@
Euler-Lagrange equations of a variational principle? For example, the %%@
first major result was by  Helmholtz in 1887. This topic also has %%@
philosophical aspects---not least the question I raised in Section %%@
5.1.2.B, about how to argue for (Eliminate) for non-simple systems. 

Though I cannot here develop this topic (cf. e.g. Santilli (1979), %%@
Lopuszanski (1999)), I should  end by considering a small aspect of it: %%@
viz. a general correspondence between sets of canonical equations and %%@
variational principles---variational principles that even allow %%@
higher-order derivatives in the Lagrangian. I say `I should consider %%@
this', because in effect, the question which has been the focus of this %%@
whole Section---`How can it be that the actual laws of motion admit a %%@
variational formulation?'---gets from this correspondence a technical %%@
interpretation---and an answer. 

The key idea is that the modified Hamilton's Principle provides a %%@
correspondence between a general class of variational problems and %%@
systems of ordinary differential equations arranged in conjugate %%@
pairs.\footnote{What follows is ``well-known'': it was discovered by %%@
Ostrogradskii writing in 1850! My summary is based on Lanczos (1986, %%@
pp. 170-72). For details and references about Ostrogradskii, cf. %%@
Kolmogorov and Yushkevich (1998, p. 201-207).} The class of variational %%@
problems is given by the extremization of an integral
\be
\int \; L(q_i, {\dot q}_i, {\ddot q}_i, \dots, q^{(m)}_i, t) \; dt \; ;
\ee
where $(m)$ indicates the $m$th derivative with respect to $t$; and %%@
where $L$ is of course an arbitrary function (it need not have %%@
mechanical significance); with the extremization subject to not only %%@
the $q_i$, but also their derivatives upto the $(m-1)$th, being %%@
prescribed at the end-points. 

I shall describe the correspondence for the simplest case beyond the %%@
already familiar one, i.e. $m = 1$,  $\dd \int  L(q_i, {\dot q}_i,t) \; %%@
dt = 0$. That is, I shall allow at most a second time derivative as an %%@
argument of $L$. I shall also assume just one degree of freedom. It %%@
will be clear enough how the correspondence generalizes to more than %%@
one $q$, and to yet higher derivatives.

Consider then the extremization of 
\be
\int \; L(q, {\dot q}, {\ddot q}, t) \; dt \; 
\ee
subject to $q$ and $\dot q$ being prescribed at the initial and final %%@
times. One easily adapts the usual calculus of variations argument to %%@
this case. The boundary  conditions now require the arbitrary function %%@
representing the variation of $q$, say $\eta$, not only to vanish at %%@
the end-points, but also to have vanishing first derivative there.  The %%@
deduction proceeds much as usual, but now includes an integration by %%@
parts of $\frac{\pl L}{\pl {\ddot q}}{\ddot {\eta}}$, as well as %%@
integrations by parts of $\frac{\pl L}{\pl {\dot q}}{\dot {\eta}}$. We %%@
get:
\be
\dd \int  L(q, {\dot q}, {\ddot q}, t) \; dt \; =  \; 0 \;\; \; \mbox{ %%@
iff } \;\;\;
\frac{\pl L}{\pl q} - \frac{d}{dt}\left( \frac{\pl L}{\pl {\dot q}} %%@
\right)  + \frac{d^2}{dt^2}\left(\frac{\pl L}{\pl {\ddot q}} \right) = %%@
\; 0.
\label{ELforLcontainssecond}
\ee

We proceed to find corresponding canonical equations. First we define a %%@
``momentum'' $u := \frac{\pl L}{\pl {\ddot q}}$, and then perform a %%@
Legendre transformation, defining $H \equiv H(q,{\dot q},u,t) := %%@
u{\ddot q} - L$; so that $L = u{\ddot q} - H(q,{\dot q},u,t)$. So our %%@
variational problem $\dd \int L \; dt = 0$ is modified to $\dd \int \; %%@
[u{\ddot q} - H(q,{\dot q},u,t)] \; dt = 0$. An integration by parts of %%@
the first term reduces this to a variational problem of the familiar %%@
kind, in $q,u$ and their first derivatives: i.e. $\dd \int \; %%@
[-H(q,{\dot q},u,t) - {\dot u}{\dot q}] \; dt = 0$. Now given {\em %%@
this} problem, we can  in the familiar way define conjugate momenta, %%@
$p_1, p_2$ say, of $q, u$, and get two pairs of canonical equations. %%@
These are {\em equivalent} to the differential equation eq. %%@
\ref{ELforLcontainssecond}.

This method easily generalizes to any number of degrees of freedom, %%@
$q_i$; and it generalizes to higher derivatives than the second. In the %%@
general case of $m$th derivatives, we first reduce them to $(m-1)$th %%@
derivatives by an integration by parts, and then repeat the process %%@
until eventually only first derivatives appear in the integrand, and we %%@
can pass to the corresponding canonical equations.  

This result also gives a characterization of the differential equations %%@
corresponding to a variational principle (of the above class). Though %%@
an arbitrary system of differential equations can be given the form
\be
{\dot q}_i = f_i(q_1,\dots,q_n,t)
\ee
by introducing suitable independent variables $q_1,\dots,q_n$, in %%@
general the functions $f_i$ will of course be different for different %%@
$i$. On the other hand:  differential equations obtained from a %%@
variational principle are derivable from a {\em single} function $H$ by %%@
differentiation. In short: Hamilton's canonical  equations are a normal %%@
form for the differential equations arising from a variational %%@
principle.

To sum up: we have shown how to pass from an arbitrary variational %%@
principle (of our class) to a system of canonical equations, all %%@
first-order in time and with all variables' time-derivatives given by %%@
differentiating a single function $H$. In effect, this result takes %%@
this Section's over-arching question---`how can it be that the actual %%@
laws of motion admit a variational formulation?'---as a technical %%@
question (instead of as a philosophical question, as in Section %%@
\ref{philosophicalpsa}). And the result, i.e. the correspondence %%@
between a large class of variational problems and sets of canonical %%@
equations, answers as follows:---\\
\indent `This is possible because the actual laws of motion, i.e. %%@
Hamilton's equations, have the very special feature that their %%@
right-hand sides, that specify the time derivatives of all the %%@
variables, are all derivatives of one and the same function $H$. If %%@
that were not so, one could {\em not} pass by a Legendre transformation %%@
to a variational formulation $\dd \int L dt = 0$.'
\\
\\
\\

{\em Acknowledgements}:---\\
It is a pleasure to thank: various audiences; Peter Holland, Chris %%@
Isham and Graeme Segal for conversations; Alex Byrne, Alan Hajek, Ned %%@
Hall, Paul Humphreys and Hugh Mellor for correspondence about Section %%@
5.1; Gerard Emch and Klaas Landsman for technical help; and Alexander %%@
Afriat, Robert Bishop, Larry Gould, Susan Sterrett, Michael %%@
St\"{o}ltzner and Paul Teller for comments on previous versions. All %%@
the remaining errors are mine!

\section{References}\label{sec:refs}
D. Armstrong (1997), {\em A World of States of Affairs}, Cambridge %%@
University Press.\\
V. Arnold (1989), {\em Mathematical Methods of Classical Mechanics}, %%@
Springer-Verlag (second edition).\\
S. Benton (1977), {\em The Hamilton-Jacobi Equation: a Global %%@
Approach}, Academic Press.\\
J. Bigelow (1988), {\em The Reality of Numbers}, Oxford University %%@
Press.\\
M. Born and E. Wolf (1999), {\em Principles of Optics}, Cambridge %%@
University Press (7th edition).\\
J. Butterfield (2003), `Solving all Problems, Postulating all States: %%@
Some Philosophical Morals of Analytical Mechanics'. In preparation.\\
J. Butterfield (2003a), `On Hamilton-Jacobi Theory: its Geometry and %%@
Relation to Pilot-wave Theory'. Submitted for inclusion in {\em Quo %%@
Vadis Quantum Mechanics?}, ed. A. Elitzur, Proceedings of a Temple %%@
University conference.\\
R. Courant and D. Hilbert (1953), {\em Methods of Mathematical %%@
Physics}, volume I, Wiley-Interscience (Wiley Classics 1989).\\
R. Courant and D. Hilbert (1962), {\em Methods of Mathematical %%@
Physics}, volume II, Wiley-Interscience (Wiley Classics 1989).\\ 
R. Dugas (1988), {\em A History of Mechanics}, Dover; reprint of a 1955 %%@
French original.\\
M. Giaquinta and S. Hildebrandt (1996), {\em The Calculus of %%@
Variations, volumes 1 and 2}, Springer-Verlag.\\
H. Goldstein et al. (2002), {\em Classical Mechanics}, %%@
Addison-Wesley.\\
W. Hamilton (1833), `On a General Method of Expressing the Paths of %%@
Light, and of the Planets, by the Coefficients of a Characteristic %%@
Function', {\em Dublin University Review}, October 1833, 795-826.\\
W. Hamilton (1834), `On the Application to Dynamics of a General %%@
Mathematical Method previously Applied to Optics', {\em British %%@
Association Report}, 1834, 513-518.\\
A. Kolmogorov and A. Yushkevich ()eds.), (1998), {\em Mathematics in %%@
the Nineteenth Century}, volume 3, Birkhauser, translated from the %%@
Russian by R. Cooke.\\
C. Lanczos (1986), {\em The Variational Principles of Mechanics}, Dover %%@
(4th edition).\\
D. Lewis (1973), {\em Counterfactuals}, Oxford: Blackwell.\\
D. Lewis (1979), `Attitudes {\em de Dicto} and {\em de Se}', {\em %%@
Philosophical Review} {\bf 88}, pp.513-543; reprinted in his (1986a), %%@
pp.133-159.\\
D. Lewis (1979a), `Counterfactual Dependence and Time's Arrow', {\em %%@
Nous} {\bf 15}, pp.455-476; reprinted in his (1986b), pp.32-66.\\
D. Lewis (1980), `A Subjectivist's Guide to Objective Chance', in R.C. %%@
Jeffrey ed. {\em Studies in Inductive Logic and Probability} volume II %%@
University of California Press; reprinted in his (1986b), pp.83-113.\\
D. Lewis (1986), {\em On the Plurality of Worlds}, Oxford: Blackwell.\\
D. Lewis (1986a), {\em Philosophical Papers, volume I}, New York: %%@
Oxford University Press.\\
D. Lewis (1986b), {\em Philosophical Papers, volume II}, New York: %%@
Oxford University Press.\\
D. Lewis (1988), `Statements partly about observation', {\em %%@
Philosophical Papers} {\bf 7}, p. 1-31; reprinted in his {\em Papers in %%@
Philosophical Logic} (1998), Cambridge University Press pp.125-155.\\
D. Lewis (1992), `Armstrong on Combinatorial Possibility', {\em %%@
Australasian Journal of Philosophy} {\bf 70}, pp.211-224; reprinted in %%@
his {\em Papers in Metaphysics and Epistemology} (1999), Cambridge %%@
University Press pp.196-214.\\
D. Lewis (1994), `Humean Supervenience Debugged', {\em Mind} {\bf 103}, %%@
p 473-490; reprinted in his {\em Papers in Metaphysics and %%@
Epistemology}, Cambridge University Press pp.224-247.\\
D. Lewis (2003), `Things qua Thruthmakers', in H.Lillehammer and G. %%@
Rodriguez-Pereyra ed.s {\em Real Metaphysics: Essays in honour of D.H. %%@
Mellor}, Routledge pp.25-42.\\
R. Littlejohn (1992), `The Van Vleck Formula, Maslov Theory and Phase %%@
Space Geometry', {\em Journal of Statistical Physics} {\bf 68}, 7-50.\\
J. Lopuszanski (1999), {\em The Inverse Variational Problem in %%@
Classical Mechanics}, World Scientific.\\
D.H. Mellor (2003), `Truthmakers', forthcoming in a volume {\em %%@
Truthmakers}, ed. H. Beebee.\\
K. Mulligan, P. Simons and B. Smith (1984), `Truth-makers', {\em %%@
Philosophy and Phenomenological Research} {\bf 44}, p. 287-321.\\
J. Papastavridis (2002), {\em Analytical Mechanics}, Oxford University %%@
Press.\\
H. Rund (1966), {\em The Hamilton-Jacobi Theory in the Calculus of %%@
Variations}, Van Nostrand.\\
R.M. Santilli (1979), {\em Foundations of Theoretical Mechanics, vol. %%@
I}, Springer-Verlag \\
M. St\"{o}ltzner (2003), `The Principle of Least Action as the Logical %%@
Empiricist's Shibboleth', forthcoming in {\em Studies in History and %%@
Philosophy of Modern Physics}.\\
E. Whittaker (1959), {\em A Treatise on the Analytical Dynamics of %%@
Particles and Rigid Bodies}, Cambridge University Press (4th %%@
edition).\\
W. Yourgrau and S. Mandelstam (1979), {\em Variational Principles in %%@
Dynamics and Quantum Theory}, Dover.

\end{document}